\documentclass[twocolumn,prl,english]{revtex4-1}
\usepackage[T1]{fontenc}
\usepackage[latin9]{inputenc}
\usepackage{verbatim}
\usepackage{amsmath}
\usepackage{amssymb}
\usepackage{graphicx}
\usepackage{babel}
\usepackage{epstopdf}
\usepackage{bigints}
\usepackage{xcolor}
\usepackage{bm}

\newcommand*\diff{\mathop{}\!\mathrm{d}}

\newcommand{\Lagr}{\mathcal{L}}

\newcommand{\SC}[1]{}
\begin{document}

\title{Nonlinear stress relaxation of transient-crosslinked biopolymer networks}
\author{Sihan Chen$^{1,2}$, Chase P. Broedersz$^{3,4}$, Tomer Markovich$^2$ and Fred C. MacKintosh$^{1,2,5,6}$}

\affiliation{$^1$Department of Physics and Astronomy, Rice University, Houston, TX 77005, USA\\
	$^2$Center for Theoretical Biological Physics, Rice University, Houston, TX 77005, USA\\
	$^3$Department of Physics and Astronomy, Vrije Universiteit Amsterdam, 1081 HV Amsterdam, The Netherlands\\
	$^4$Arnold-Sommerfeld-Center for Theoretical Physics and Center for NanoScience, Ludwig-Maximilians-Universit\"at M\"unchen, D-80333 M\"unchen, Germany\\
	$^5$Department of Chemical and Biomolecular Engineering, Rice University, Houston, TX 77005, USA\\
	$^6$Department of Chemistry, Rice University, Houston, TX 77005, USA}

\begin{abstract}
		 A long standing puzzle in the rheology of living cells is the origin of the experimentally observed long time stress relaxation. The mechanics of the cell is largely dictated by the cytoskeleton, which is a biopolymer network consisting of transient crosslinkers, allowing for stress relaxation over time. Moreover, these networks are internally stressed due to the presence of molecular motors. In this work we propose a theoretical model that uses a mode-dependent mobility to describe the stress relaxation of such prestressed transient networks. Our theoretical predictions agree favorably with experimental data of reconstituted cytoskeletal networks \SC{and may provide an explanation for the slow stress relaxation observed in cells. }
\end{abstract}
\maketitle

\section{Introduction}

Living cells are known to exhibit unusual mechanical properties including an internal nonlinear stiffening under external stress, in which their stiffness can increase by orders of magnitudes \cite{Trepat2007592}, reversible softening under compression~\cite{Chaudhuri2007}, viscoplasticity~\cite{Bonakdar2016} and poroelasticity~\cite{Kimpton2015}. A long-standing puzzle is related to the surprisingly slow stress relaxation that has been measured in living cells\SC{~\cite{Navarro1996, Fabry2001,Stamenovic2004,Desprat20052224,Bursac2005557,Balland2006,Stamenovic2007}}. 
This stress relaxation reveals more than just a long relaxation time, but also a broad spectrum of relaxation times, with
a dynamic modulus that varies with frequency as a power-law with exponent $\beta$ in the range of $\sim0.1- 0.3$~\cite{Fabry2001}.
It has been argued that this may be related to the soft glassy rheology (SGR) model \cite{Sollich1998,Fabry2001,Bursac2005557}, although the relevance and validity of this in cell mechanics remains unclear. 

Most of the mechanical properties of living cells originate in the cytoskeleton, a dynamic network composed of crosslinked biopolymers, which gives the cell its shape and rigidity~\cite{Hall1998}. One reason for its dynamic nature is that many of the crosslinking proteins, for example $\alpha$-actinin, form transient bonds with both finite binding and unbinding rates~\cite{Schiffhauer20161473}. Such crosslinking proteins, denoted as transient crosslinkers, introduce a distinct type of stress relaxation in semiflexible polymer networks, since the unbinding of crosslinkers allows the networks to flow at long time~\cite{Ward20084915,Lieleg2008,Lieleg20094725,Broedersz2010,Yao2013,Muller2014}. Previous theory and experiments involving reconstituted biopolymer networks have revealed a characteristic scaling exponent of $\beta=1/2$ for the frequency-dependent linear shear modulus~\cite{Broedersz2010,Yao2013,Muller2014}. Unlike the Rouse model, in which the same exponent appears in the high-frequency regime~\cite{RouseJr.19531272,gennes_1990,doi}, transient networks only exhibit the $1/2$ exponent in the low-frequency regime, indicating a different mechanism of stress relaxation from the Rouse model. Moreover, the Rouse model applies to flexible polymers, while the analogous high-frequency regime for semiflexible polymers such as actin is known theoretically and experimentally to exhibit a $3/4$ exponent~\cite{Gittes1997,Gittes1998,Morse1998, Morse19987030,Koenderink2006,Broedersz2014, Pritchard2014}. 

Within the cytoskeleton there are also molecular motors that generate internal stresses~\cite{Alberts,Bray2001,Markovich2019}, which may alter the rheological properties of living cells~\cite{Wang2002, Stamenovic2004, FERNANDEZ2006}.  Recently, it has been shown experimentally that the apparent scaling exponent of the linear shear modulus in reconstituted actin networks with transient crosslinks can be further reduced and systematically varied over the range of $0.1\lesssim\beta\le0.5$ by an applied external stress~\cite{Mulla2019}. 
\SC{It is well known that applying external or internal stress on a permanent biopolymer networks can cause nonlinear stiffening~\cite{Gardel1301,Storm2005191,Mizuno2007,Koenderink2009} and a reduction in the high-frequency exponent from 3/4 to 1/2~\cite{ Granek1997,Caspi1998,Rosenblatt2006, Mizuno2007,Majumdar2008}}. When applied to transient networks, aside from the stress-stiffening response, external stress  solidifies the network~\cite{Yao2013}, suppressing both stress relaxation and reducing the apparent exponent of the frequency-dependent nonlinear shear modulus~\cite{Mulla2019}. This provides a possible explanation for the weak scaling exponent observed in living cells, as living cells are intrinsically under internal stress generated by molecular motors or other active processes~\cite{Alberts,Julicher20073,Roh-Johnson20121232, Chen2020}, although a microscopic model for how stress qualitatively changes the dynamics is still lacking.

In this paper, we develop such microscopic theory for the mechanical response of transient-crosslinked semiflexible networks under applied stress. Based on equilibrium thermodynamics, we show that the dynamics of transiently crosslinked semiflexible polymer networks can be described by a mode-dependent mobility. We analytically derive the form of this mobility in transient networks, and calculate the nonlinear modulus under external stress. Our theoretical prediction of the nonlinear modulus quantitatively agrees with experiment data of reconstituted cytoskeletal networks. We also show that external stress naturally leads to a weak frequency dependence by suppressing fluctuations of bending modes. 

\begin{figure}[h]
	\centering
	\includegraphics[scale=0.35]{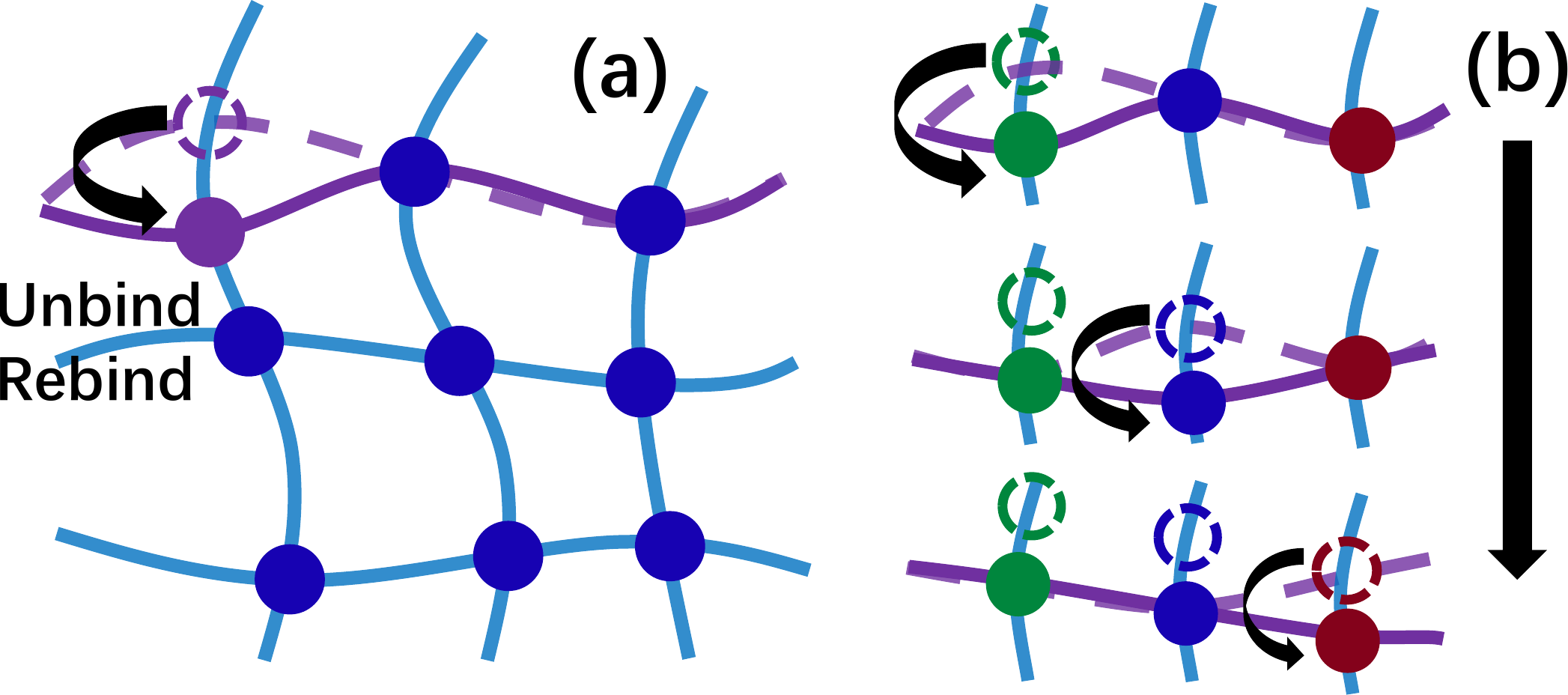}
	\caption{(a) Schematic figure of a transient network, \SC{in which polymers (lines) are connected by transient crosslinkers (circles). The unbinding and rebinding of a single crosslinker (purple circle) relaxes its adjacent segments.} (b) The relaxation of a long-wavelength mode requires successive unbinding events (\SC{green circle -> blue circle -> brown circle in sequence }). }
	\label{Fig.1}
\end{figure}

\section{Overview}
\SC{Transient crosslinkers introduce a distinct type of stress relaxation: when a crosslinker unbinds, it relaxes the stress on adjacent polymer segments (the crosslinker later bind on a different position), see Fig.~\ref{Fig.1}(a). In order for a long polymer to relax, multiple successive unbinding events are required (Fig.~\ref{Fig.1}(b)), resulting in a relaxation time much longer than the timescale of a single unbinding event~\cite{Broedersz2010,Broedersz2014}. }

\begin{figure}[t]
	\centering
	\includegraphics[scale=0.4]{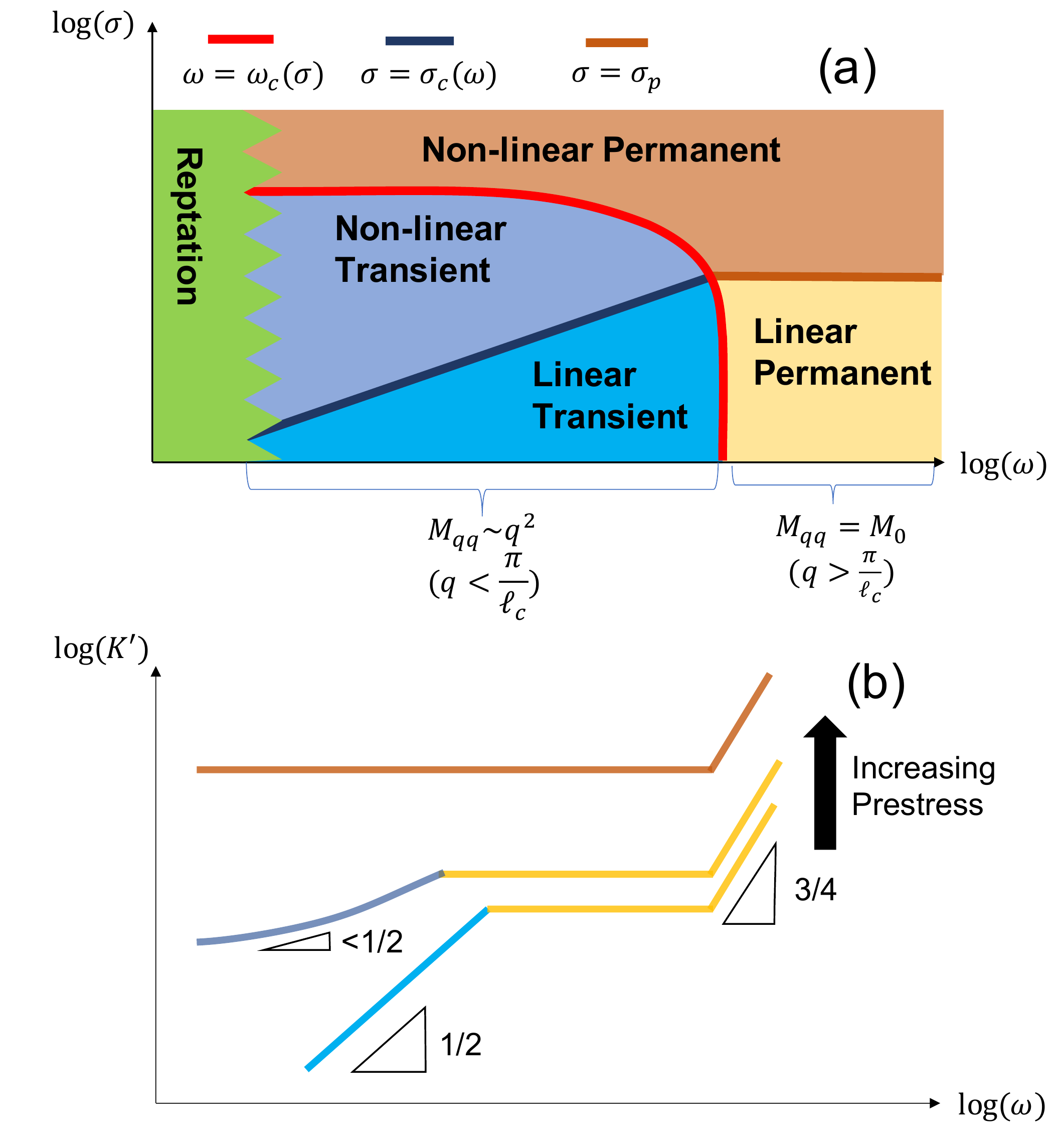}
	\caption{(a) Schematic regime diagram of a transient network as function of frequency $\omega$ and prestress $\sigma$. The extremely low-frequency regime is dominated by reptation, which is not considered in our model. The rest of the diagram consists of a transient and permanent regimes, separated by $\omega_c(\sigma)$, which is the characteristic frequency, see Eq.~(\ref{e37}).  $\sigma_c$ and $\sigma_p$ are the characteristic stresses for the nonlinear-stiffening in the transient and permanent regimes, respectively. (b) Schematic curves of \SC{the differential elastic modulus $K'$} versus $\omega$, for small prestress (linear transient and linear permanent), intermediate prestress (non-linear transient and linear permanent) and large prestress (nonlinear permanent).  The corresponding power-law regimes are indicated. With stress, the low-frequency regime is not predicted to be a strict power law, although we show that it may appear to be so. }
	\label{Fig.rgm}
\end{figure}

\SC{We propose a microscopic model that accounts for the effect of unbinding/rebinding of transient crosslinkers.} In our model, that is described in detail below, we show that the relaxation of semiflexible polymer networks can be decomposed into the relaxation of the independent bending modes (see Sec.~\ref{section:3}). The relaxation of each mode follows a mode-dependent mobility, $M_{qq}$, \SC{($q$ being the wave number)}  which leads to non-trivial dependences on the frequency of the linear and nonlinear viscoelastic moduli. 
In Fig.~\ref{Fig.rgm} (a) we show a schematic diagram of the various regimes  for a transient-crosslinked biopolymer network in the ($\omega,\sigma$) phase-space, where $\omega$ is the frequency and $\sigma$ is the applied shear stress. With only transient crosslinkers in the system, reptation of finite molecular weight polymers is expected to lead to liquid-like behavior on the longest timescales. In the present model we have focused on stress relaxation that is entirely due to transient-crosslinking, and we therefore consider the limit of high molecular weight and timescales less than the reptation time. On these timescales, we find that the stress relaxation can be devided into two regimes: in the low-frequency regime the stress relaxation is governed by the transient behavior of the crosslinkers, while in the high-frequency regime the network behaves as if the crosslinkers were permanent~\cite{Broedersz2010}. The two regimes are separated by a characteristic frequency, $\omega_c(\sigma)$, which depends on $\sigma$ (see Sec.~\ref{section:5}, Eq.~(\ref{e37})).  These two different regimes appear as a result of the different $q$-dependence of the mode-dependent mobility $M_{qq}$, for $q>\pi/\ell_c$ and $q<\pi/\ell_c$, where $\ell_c$ is the average spacing between crosslinkers in the network. The relaxation of bending modes with $q>\pi/\ell_c$ is dominated by the solvent viscosity, as in permanent networks~\cite{Gittes1998, Morse1998}. However, the relaxation of bending modes with $q<\pi/\ell_c$ is limited by the transient nature of the  crosslinkers, leading to $M_{qq}\sim q^2$ (see Sec.~\ref{section:4}). This quadratic dependence in $M_{qq}$ results in a linear modulus $G(\omega) \sim \omega^{1/2}$. Moreover,  when  external stress is applied, the network may stiffen nonlinearly, where the characteristic stress, $\sigma_c(\omega)\sim \omega^{1/3}$, governs the transition from the linear to the nonlinear stiffening regimes (see Sec.~\ref{section:5}, Eq.~(\ref{e22})). When the network stiffens nonlinearly, \SC{the differential shear modulus $K=K'+iK''=\diff \sigma/\diff \gamma $ is used to characterize the viscoelastic behavior, where $\gamma$ is the shear deformation. }~\SC{As we show below (see Sec.~\ref{section:6}) and sketched Fig.~\ref{Fig.rgm}(b), the apparent exponent with which $G$ depends on $\omega$ is reduced with applied stress and can become arbitrarily small at high stress. This is consistent with recent experiments~\cite{Mulla2019} and may provide an explanation for the week dependence seen in living cells~\cite{Navarro1996, Fabry2001,Stamenovic2004,Desprat20052224,Bursac2005557,Balland2006,Stamenovic2007}.} 
\section{Model}
\label{section:3}
To predict the mechanical response of a biopolymer network under stress, we begin by describing a single semiflexible polymer under tension. For simplicity we first discuss a polymer moving on a 2D plain where the transverse deformation is limited in one direction, and then extend the result to polymers in 3D. The Hamiltonian for such polymers with length $\ell$ under tension $F$ is~\cite{Gittes1998,Morse1998,Broedersz2014}:

\begin{equation}
\begin{aligned}
H=\frac \kappa 2 \int dx \,\left(\frac {\partial^2u}{\partial x^2}\right)^2 + \frac F 2 \int \diff x \left(\frac{\partial u}{\partial x}\right)^2 .
\end{aligned}
\label{e1}
\end{equation}
Here $\kappa$ is bending rigidity and $u(x)$ is the transverse displacement at position $x$. The first term in Eq.~(\ref{e1}) is the bending energy of the chain, while the second term is the work done by the external force, where $\Delta \ell = \int \diff x (\partial u/\partial x)^2/2$ is the contraction of the end-to-end distance due to bending fluctuations. This transverse displacement can be decomposed using Fourier series to a series of bending modes, $\{u_q\}$, with $q$ being the wave number:
\begin{equation}
\begin{aligned}
u(x)=\sqrt{\frac{ 2}{ \ell}}\sum_q u_q {\sin}(qx) \qquad \left(q=\frac{n\pi}\ell\right) ,
\end{aligned}
\label{e2}
\end{equation}
where the dimension of $u(x)$ is set to be $[L]^{1/ 2}$ to simplify the expression of the Hamiltonian. Using Eq.~(\ref{e2}), the Hamiltonian in Eq.~(\ref{e1}) is diagonalized and assumes a simple quadratic form: 
\begin{equation}
\begin{aligned}
H= \frac{1}{2}\sum_q\left(\kappa q^4+F q^2\right)u_q^2.
\end{aligned}
\label{e3}
\end{equation}
The dynamics of the amplitudes of all bending modes follow a standard Langevin equation with $\{u_q\}$ as variables (this model is usually referred to as Model A) \cite{Hohenberg1977435}, 

\begin{align}
&\frac{du_q}{dt}=-\sum_p M_{qp}(\{u_s\})\frac{\partial H}{\partial u_p}+\eta_q\notag
\\&=-\sum_p M_{qp}(\{u_s\}) \left(\kappa p^4+F p^2\right)u_p+\eta_q,\label{e4}
\end{align}
where $M_{qp}(\{u_s\})$ is the generalized mobility matrix, $\{u_s\}$ denotes the set of amplitudes of all bending modes, \{$u_{\pi/\ell}$, $u_{2\pi/\ell}$, ...\}, and $\eta_q$ is a thermal Gaussian white noise with zero mean and variance  $\langle \eta_p(t)\eta_q(t')\rangle=2k_BT M_{pq}\delta(t-t')$, where $k_B$ is  Boltzmann constant and $T$ is the temperature.

In general, $M_{pq}$ is a function of all $\{u_s\}$, and Eq.~(\ref{e4}) is nonlinear. However, the transverse displacement of semiflexible polymers is small ($u(x) \ll \ell$), and as we shall see below, when applying an external stress these displacements are even smaller. Therefore, within our framework, Eq.~(\ref{e4}) should be linearized, leading to a constant mobility matrix $M_{pq}$. The evolution of the correlation function of mode $q$, $\langle u_q(0)u_q(t)\rangle$, is then

\begin{align}
&\frac{d}{dt}\langle u_q(0)u_q(t)\rangle=\left\langle u_q(0)\frac{du_q(t)}{dt}\right\rangle\notag\\&=-\sum_pM_{pq} \left(\kappa p^4+F p^2\right)\langle u_q(0)u_p(t)\rangle\notag\\&=-M_{qq} \left(\kappa q^4+F q^2\right)\langle u_q(0)u_q(t)\rangle,\label{e5}
\end{align}
where $\langle ... \rangle$ denotes average over thermal noise realizations. In the last equation we use the fact that $H(\{u_q\})$ is diagonal, implying that there are no correlations between different bending modes, i.e. $\langle u_p(0)u_q(t) \rangle \sim \delta_{pq}$. Thus, the correlation function follows a simple exponential decay, where the variance at $t=0$ is obtained from Eq.~(\ref{e3}) using the equipartition theorem (assuming the system is in equilibrium at $t=0$)~\cite{Gittes1998}, such that :
\begin{equation}
\begin{aligned}
\langle u_q(0)u_q(t)\rangle=\frac{k_B T}{\kappa q^4+F q^2}e^{-M_{qq}({\kappa q^4+F q^2})t}.
\end{aligned}
\label{e6}
\end{equation}
For every bending mode $q$, there is only one parameter associated with its relaxation process, $M_{qq}$, which is the mobility for mode $q$. This mode-dependent mobility naturally emerges  from the linearization of the Langevin equation and can thus be generally applied to any  semiflexible polymer networks. Once the mode-dependent mobility is known, the dynamics of the network is determined. 

Notice that Eq.~(\ref{e6}) shows that the variance of $u_q$ decreases with increasing $F$, thus tension reduces bending fluctuations. This is consistent with our assumption that linearizing Eq.~(\ref{e4}) is always valid for semiflexible polymers. For flexible polymers, although their Hamiltonian can also be diagonalized to a quadratic form (Rouse Model~\cite{RouseJr.19531272}), the amplitude of the transverse fluctuations is large and the linearization of Eq.~(\ref{e4}) is not generally valid. 

Next, we use the correlation function of Eq.~(\ref{e6}) to calculate the correlation function of the end-to-end distance under tension $F$, $C_F(t)\equiv\langle \delta \ell(0)\delta \ell(t)\rangle$. Here $\delta \ell = \langle \Delta \ell \rangle - \Delta \ell$ is the projected end-to-end extension of the polymer with respect to its rest length  for small $u$ and $\Delta \ell = \int \diff x (\partial u/\partial x)^2/2$. This relation leads to a simple formula for $C_F(t)$~\cite{Gittes1998}:

\begin{align}
&C_F(t)= \sum_q q^4\langle u_q(0)u_q(t)\rangle^2\notag\\
&=\sum_q \frac{(k_B T)^2 \,q^4}{\left(\kappa q^4+F \,q^2\right)^2} \exp{\left[-2M_{qq} {\left(\kappa q^4+ F q^2\right)}t\right]}. \label{e8}
\end{align}
The Fourier-Transform of the correlation function is then used to compute the end-to-end power spectrum, $P_F(\omega)\equiv |\delta \ell(\omega)|^2$,

\begin{equation}
\begin{aligned}
P_F(\omega)=\sum_q \frac{(k_B T)^2 \,q^4}{\left(\kappa q^4+F \,q^2\right)} \frac{4M_{qq} }{\omega^2+4M_{qq}^2 {\left(\kappa q^4+ F q^2\right)}^2}. 
\end{aligned}
\label{e27}
\end{equation}
Using the fluctuation-dissipation theorem, we can relate this power spectrum to the response function $\chi(\omega;F)$: $\ell \chi''(\omega;F)=\omega P_F(\omega)/2k_B T$, which together with the  Kramers-Kronig relations gives
\begin{equation}
\begin{aligned}
\!\!\!\!\chi(\omega;F) &\!= \! \sum_q\!  \frac{2k_B TM_{qq}\left({\kappa q^4+F q^2}\right)  }  {\ell\left(\kappa q^2+F\right)^2\left[2M_{qq}{\left(\kappa q^4+F q^2\right)} -i\omega\right] }.\!
\end{aligned}
\label{e18}
\end{equation}
This response function describes the mechanical response of the chain, given that the tension is slightly perturbed around $F$. For polymers in 3D, one should add a factor of two to the right-hand-side of Eq.~(\ref{e18}) in order to account for the transverse displacement in two directions. In the rest of the paper we will use the 3D result. As we will show later, the macroscopic modulus of the entire network can be derived using the response function of a single polymer, Eq.~(\ref{e18}). 

\section{Mode-dependent Mobility }
\label{section:4}
\subsection{Mode-dependent Mobility for Transient-crosslinked Networks}

In the previous section we have shown that the dynamics of a single polymer in any biopolymer network is well described by a mode-dependent mobility $M_{qq}$, while its specific form depends on  the network structure. In this section we first derive $M_{qq}$ for transient networks in the small-$q$ limit in a simple and intuitive way, and then detail a general method that can be used to derive $M_{qq}$ for any $q$.  

Lets us consider a transient netwrok in the hydrodynamic limit (i.e. long-wavelength limit). For small $q$, we can Taylor expand $M_{qq}$,
\begin{equation}
\begin{aligned}
M_{qq} = a_0+a_2 q^2 +a_4 q^4...,
\end{aligned}
\label{e24}
\end{equation}
where $a_n$ denotes the coefficient of the $n$-degree term. The form of $M_{qq}$ is constrained by polymer symmetries. Therefore, since the polymer does not have a preferred transverse direction (it can have a preferred longitudinal direction though due to the polarity~\cite{Alberts}) $M_{qq}$ must be an even function of $q$. The term $a_0$ stands for the mobility for the $q=0$ mode (infinite wavelength). For transient networks, the value of $a_0$ must be $0$. The reason for this can be seen from Fig.~\ref{Fig.1}(b), where we sketch the relaxation process of a long-wavelength mode. In order to relax such a mode, multiple successive unbinding events are required, indicating that transient crosslinkers impose stronger limitations to long-wavelength modes, leading to smaller mobilities. In order to relax the infinite-wavelength mode there should be infinite successive binding events, each of them takes finite time, therefore leading to zero mobility. Hence, the leading term in Eq.~(\ref{e24}) is the quadratic term, and for small $q$ we have $M_{qq}=a_2 q^2$. 

So far we have shown that the transient nature of the network results in a quadratic dependence of $M_{qq}$. However, this quadratic dependence should only be valid for $q\ll q_c$, where $q_c = \pi/\ell_c$ is the characteristic wave number with wavelength $\ell_c$. 
Bending modes with $q\gg q_c$ are not limited by transient crosslinking, and their relaxation is determined by the substrate mode-independent viscosity (rather than the networks itself), which corresponds to a constant mobility $M_0$. Together, we can approximate $M_{qq}$ for all wavelengths:
\begin{equation}
\begin{aligned}
M_{qq}=
\left\{ 
\begin{aligned}
&{a_2 q^2} \qquad&(q \leq q_c )\\
&M_0  \qquad&(q >q_c)
\end{aligned}
\right.. 
\end{aligned}
\label{e16}
\end{equation}
This mobility is discontinuous at $q=q_c$, because we simply separate the bending modes into a crosslink-limited and viscous-dominated parts. In fact,
for bending modes with $q\sim q_c$, both the transient crosslinkers and the substrate viscosity contribute to the stress relaxation, and we anticipate a transition in $M_{qq} $ from the quadratic dependence to the constant mobility. We determine the dependence of $M_{qq}$ for $q\sim q_c$  in Sec.~\ref{section:4}.~B.

Although this form of the mobility is sufficient for predicting the macroscopic modulus (see Sec.~\ref{section:5}), a microscopic understanding of the parameter $a_2$ is important for understanding the physics of the model we use. In general, the mobility (and thus also $a_2$) can depend on $F$, for example due to catch/slip bond. For simplicity we neglect this effect hereafter (see Sec.~\ref{section:6} for further discussion). We can then consider the dynamics of a polymer for $F=0$ and calculate $a_2$. The relaxation time of bending mode $q$, $\tau_q$, can be read from Eq.~(\ref{e6}). When $F=0$, it is $\tau_q = 1/(\kappa q^4 M_{qq})$. For small $q$, we have $\tau_q = 1/(a_2 \,\kappa q^6)$. On the other hand, the relaxation process for small $q$ is limited by binding/unbinding of crosslinkers, which is characterized by a single timescale $\tau_{\rm off}$ (the unbinding time, as appropriate for strong crosslinkers that spend most of their time in the bound state~\cite{Broedersz2010}), therefore, we  also have $\tau_q \sim \tau_{\rm off}$. From dimensional analysis, the coefficient $a_2$ can be written in terms of microscopic parameters:
\begin{equation}
\begin{aligned}
a_2 = c\, \frac{\ell_c^6}{\kappa \tau_{\rm off}},
\end{aligned}
\label{e31}
\end{equation}
where $c$ is a dimensionless constant. Here $\ell_c$ is the average spacing between crosslinkers which appears as it is the only lengthscale associated with transient crosslinkers. 

In order to find the value of $c$, we use the (mean-field) result of Ref.~{\cite{Broedersz2010}} for the response function $\chi$ for $F=0$ (see Appendix A for a complete derivation). In this case, the relaxation rate of each bending mode, $\omega_r(q)$, can be written as \cite{Gittes1998}
\begin{equation}
\begin{aligned}
\omega_r(q) = \frac{2\kappa q^4}{\xi(q)}.
\end{aligned}
\label{e9}
\end{equation}
where $\xi(q)=1/M_{qq}$ is the mode-dependent friction. Since a mode with longer wavelength must have longer relaxation time, $\omega_r(q)$ must increase monotonically, suggesting the existence of the inverse function $q(\omega_r)$. The slowest relaxation rate is $\omega_r(q=\pi/\ell)$ ($\ell$ being the polymer length), corresponding to the longest wavelength mode, which must vanish as $\ell\to\infty$. Therefore, in the long-chain limit we always have $\omega \gg \omega_r(q=\pi/\ell)$. This allows us to approximate the summation in Eq.~(\ref{e18}) with an integral (as will be done in the rest of the paper), leading to:

\begin{eqnarray}
\chi(\omega;F=0)&\simeq& \frac{2k_B T}{\pi\kappa}\int_0^{\infty} \diff q \, \frac{2a_2{ q^2}  }  { \omega_r(q) -i\omega} \nonumber\\&=&\frac{a_2^{1/ 2}k_B T}{3\kappa^{3/ 2	}}\omega^{-1/2}(1+i),\label{e26}
\end{eqnarray}
where the $-1/2$ exponent is consistent with Ref.~\cite{Broedersz2010}. In Ref.~\cite{Broedersz2010}, the linear response function $\chi(\omega;F=0)$ is derived using a mean-field theory (see appendix for details-- note the factor of two for a polymer in 3D):
\begin{align}
\chi(\omega;F=0)&=0.0036\frac{2k_BT\ell_c^3}{\pi \kappa^2}\int_{-\infty}^{\infty} \,\frac{\diff q}{q^2-2i\omega \tau_{\rm off}}\notag\\&=0.0036\frac{k_BT\ell_c^3}{\tau_{\rm off}^{ 1/ 2} \kappa^2}\omega^{- 1/ 2}(1+i). \label{e28}
\end{align}
Comparing this response function with our result (Eq.~(\ref{e26})) gives the analytical expression of the coefficient $a_2$
\begin{equation}
\begin{aligned}
a_2 =  0.00012 \frac{\ell_c^6}{\kappa \tau_{\rm off}}. 
\end{aligned}
\label{e30}
\end{equation}

Although $\ell_c$ and $\kappa$ can be measured experimentally, the unbinding time $\tau_{\rm off}$ of the crosslinker is usually unknown. Therefore, replacing $a_2$ with Eq.~(\ref{e30}) does not reduce the number of fitting parameters. However, Eq.~(\ref{e30}) gives a microscopic understanding of the coefficient $a_2$, and can be further used to calculate $\tau_{\rm off}$ (see Sec.~\ref{section:5}).

\subsection{Mode-dependent Mobility for Generic Networks}
Heretofore we have derived the mode-dependent mobility for transient networks. We find that a quadratic dependence in the small-$q$-limit naturally emerges as a result of the transient nature of the network. However, the mobility for $q\sim q_c$ remains unclear, and the same derivation does not apply to other networks, despite the fact that the mode-dependent mobility can be applied to any semiflexible polymer networks. In this subsection we provide a general method for deriving the mode-dependent mobility for any network, provided that the correlation function of the end-to-end distance is known. The latter can be found analytically, using numerical simulations, or measured experimentally. We then use this method to derive the mode-dependent mobility of transient networks for $q\sim q_c$.

\begin{figure}[t]
	\centering
	\includegraphics[scale=0.5]{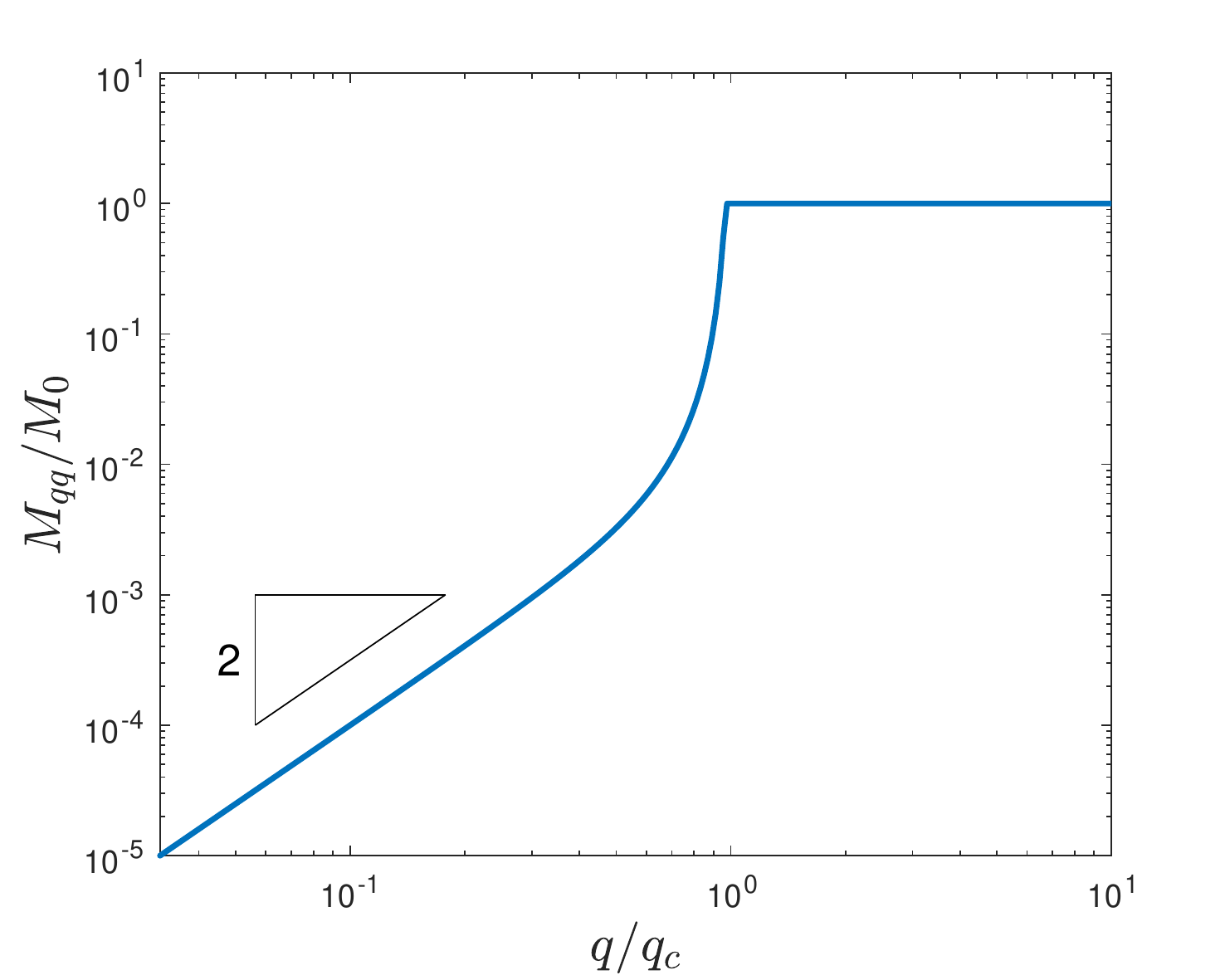}
	\caption{Mode-dependent mobility, $M_{qq}$, as a function of $q$, calculated from Eq.~(\ref{e34}) with $M_0=100a_2\kappa q_c^4$. The form of the mobility (including the cusp) is further discussed in Appendix B. }
	\label{Fig.mdm}
\end{figure}
The correlation function, $C_F(t)$, is related to the mode-dependent-mobility, $M_{qq}$, via Eq.~(\ref{e8}). Once $C_F(t)$ of a given network is obtained, $M_{qq}$ can be calculated by inverting Eq.~(\ref{e8}). For simplicity, and as the mobility is assumed to be independent of $F$, we only consider $F=0$. Following the same reasoning as in the previous section (see paragraph before Eq.~(\ref{e26})) we approximate the summation in Eq.~(\ref{e8}) with an integral:
\begin{align}
C_0(t)\equiv C_{F=0}(t)&= \frac{2\ell (k_B T)^2}{\pi \kappa^2}\int \frac{\diff q}{q^4}  \,\exp{\left[-\omega_r(q)t\right]}\notag
\\&\!\!\!\!\!\!\!\!=\frac{2\ell (k_B T)^2}{\pi \kappa^2}\int \frac{\diff \omega_r}{q^4}  \,\frac{\diff q}{\diff \omega r}\exp{\left[-\omega_r t\right]}. \label{e10}
\end{align}
Surprisingly, we find that the correlation function at vanishing force is proportional to the Laplace transform of $q^{-4}\diff q/\diff \omega_r$. The inverse Laplace transform of Eq.~(\ref{e10}) gives a differential equation for $q(\omega_r)$,
\begin{equation}
\begin{aligned}
\Lagr^{-1} \{C_0\}(\omega_r)
=\frac{2\ell (k_B T)^2}{\pi \kappa^2q^4}\,\frac{\diff q}{\diff \omega r},
\end{aligned}
\label{e11}
\end{equation}
where 
\begin{equation}
\begin{aligned}
\Lagr^{-1} \{C_0\}(\omega_r)
=\frac{1}{2\pi i}\lim_{T\to \infty}\int_{\gamma -i T}^{\gamma+i T}e^{s\omega_r}C_0(s) \diff s. 
\end{aligned}
\label{e12}
\end{equation}
Here $\gamma $ is any real number greater than the real part of all singularities of $C_0(s)$. Using Eq.~(\ref{e11}) and the correlation function $C_0(t)$, one can derive the relaxation rate $\omega_r(q)$. Then the mode-dependent friction is easily found using Eq.~(\ref{e9}).

We now use this method to derive the mode-dependent mobility of transient networks for all $q$'s. As discussed above, the stress relaxation of transient networks is governed by two different mechanisms: a slow relaxation relying on the transient crosslinkers and a fast relaxation dominated by the substrate viscosity. We assume the timescales for the two processes to be separated, i.e. $\tau_{\rm off}\gg \tau_{\rm per}$ or $a_2 \ll2 M_0 q_c^4$, where $\tau_{\rm per} = 1/(2\kappa M_0 q_c^4)$ is the longest relaxation time governed by the substrate viscosity.  Under this assumption, the contribution to the correlation function from the two relaxation processes is additive \cite{Broedersz2010}:
\begin{align}
C_0(t)&=&\frac{2(k_BT)^2\ell}{\pi \kappa^2}\int_{0}^{\infty}\frac{\diff q}{q^4} \,\exp{\left(-2a_2 {\kappa q^6}t\right)}\notag\\&\,\,+&\frac{2(k_BT)^2\ell}{\pi \kappa^2}\int_{q_c}^{\infty}\frac{\diff q}{q^4} \,\exp{\left(-2M_0 {\kappa q^4}t\right)}, \label{e13}
\end{align}
where the first term is the contribution due to the transient nature of the crosslinkers. The second term is the classic result of the correlation function of a semiflexible polymer with mobility $M_0$ where the average crosslinking distance is $\ell_c$~\cite{Gittes1998}. 
The inverse Laplace transform of Eq.~(\ref{e13}) is:
\begin{align}
& \Lagr^{-1}\{C_0\}(\omega_r)=\frac{2(k_BT)^2\ell}{\pi \kappa^2}\int_{0}^{\infty}\frac{\diff q}{q^4} \,\delta{\left(\omega_r-2a_2 {\kappa q^6}\right)}\notag\\ &+\frac{2(k_BT)^2\ell}{\pi \kappa^2}\int_{q_c}^{\infty}\frac{\diff q}{q^4} \,\delta{\left(\omega_r-2M_0 {\kappa q^4}\right)}\notag
\\& =\frac{(2a_2)^{ 1/ 2}(k_BT)^2\ell}{3\pi \kappa^{ 3/ 2}}\omega_r^{-3/ 2}\notag\\&+\frac{(2M_0)^{ 3/ 4}(k_BT)^2\ell}{2\pi \kappa^{ 5/ 4}}\omega_r^{-7/ 4}\Theta\left(\omega_r-2M_0 \kappa q_c^4\right), \label{e14}
\end{align}
where $\Theta(x)$ is the heaviside function.
Substituting Eq.~(\ref{e14}) into Eq.~(\ref{e11}) and integrating from $q$ to infinity, with the boundary condition $\omega_r(q\to \infty)=\infty$ gives:
\begin{align}
&\!\!\!\!\!(2a_2\kappa)^{1/2}\omega_r^{-1/2}+{(2M_0\kappa)^{ 3/ 4}}\omega_r^{-3/ 4}\Theta\left(\omega_r-2M_0 \kappa q_c^4\right)\notag\\&\!\!\!\!\!=q^{-3}-q_c^{-3}\Theta\left(-\omega_r+2M_0 \kappa q_c^4\right).\label{e15}
\end{align}

Because we assume $a_2\ll(2M_0 \kappa q_c^4)$, the solution of  Eq.~({\ref{e15}) is
\begin{equation}
\begin{aligned}
\omega_r(q)=
\left\{ 
\begin{aligned}
&{2 a_2\kappa}q^{6}{\left[1-\left(\frac{q}{q_c}\right)^{3}\right]^{-2}} \!\!&(q \leq q_c-\Delta q )\\
&2M_0\kappa q^4  \!\!&(q >q_c-\Delta q ),
\end{aligned}
\right.
\end{aligned}
\label{e32}
\end{equation}
where $\Delta q = {{a_2}^{ 1/ 2}q_c^2}/({3M_0^{1/ 2}})$. 
The mobility $M_{qq}$ can then be found using Eq.~(\ref{e9}),

\begin{align}
M_{qq}=
\left\{ 
\begin{aligned}
&{a_2 q^2}{\left[1-\left(\frac{q}{q_c}\right)^{3}\right]^{-2}} &(q \leq q_c-\Delta q )\\
&M_0  &(q >q_c-\Delta q).
\end{aligned}
\label{e34}
\right.
\end{align}
In Fig.~\ref{Fig.mdm} we plot the mobility $M_{qq}$ as a function of $q$. For $q\ll q_c$, $M_{qq} $ shows a quadratic dependence on $q$, in agreement with our previous analysis for the small $q$ limit. When $q$ approaches $q_c$, $M_{qq}$ increases dramatically until it reaches $M_0$. Although this cusp in $M_{qq}$ near $q=q_c$ appears unphysical, it is essential within our model with a single lengthscale $\ell_c$ for the appearance of the plateau in the linear modulus~\cite{Broedersz2010} (see Appendix B for a mathematical proof).

\begin{figure}[t]
	\centering
	\includegraphics[scale=0.41]{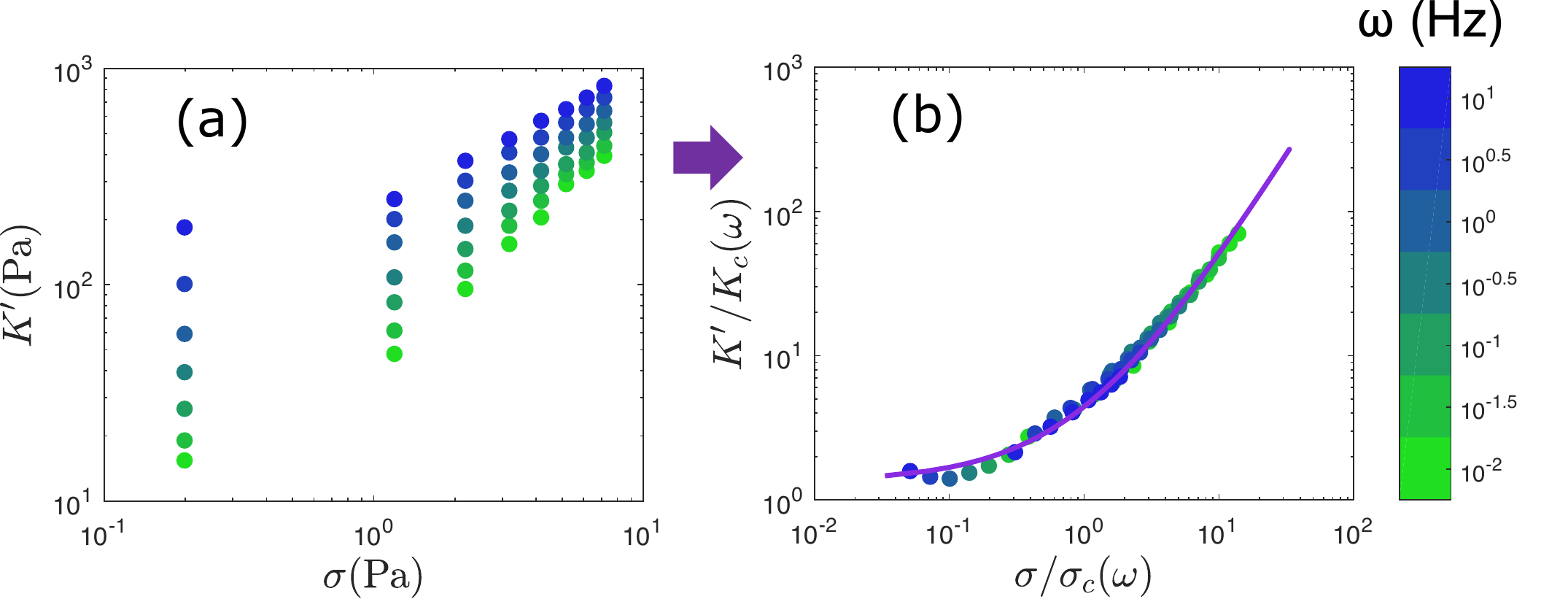}
	\caption{(a) A plot of $K'$ as function of the external stress $\sigma$ for different frequencies using experimental data of reconstituted actin networks (taken from Ref.~\cite{Mulla2019}). In (b) we plot the rescaled $K'$, vs the rescaled prestress for the various frequencies. The purple curve is the theoretical prediction of Eq.~(\ref{e20}). The best fit values are $\gamma_1 = 1.71 \,{\rm Pa \, s^{1 /3}}$, $\gamma_2 = 35.9\,{\rm Pa \, s^{ 1/ 2}}$. (fit was done using only the $K'$ data.)}
	\label{Fig.2}
\end{figure}

\section{Dynamic Modulus} 
\label{section:5}
In this section we derive the dynamic modulus of the transient network using the mode-dependent mobility calculated above. We focus on the transient nature of the network and neglect the detailed network structure by assuming all the polymers are aligned in the same direction, i.e.,~an effective  1D network. A more complete discussion of the 3D case is deferred to future publication~\cite{unpublished}. When an external prestress $\sigma$ is applied on an aligned network, all the polymers feel the same tension, $F=\sigma/\rho$, where $\rho$ is the polymer length per unit volume~\cite{Broedersz2014}. When  $\sigma$ is perturbed by $\diff\sigma$, the tension is also perturbed, $\diff F =\diff\sigma/\rho $, leading to an extension, $\diff \ell = \ell \chi \diff F$. Therefore, the perturbation in stress causes a perturbation in strain, $\diff \gamma = \diff \ell/\ell= \diff \sigma\chi /\rho$, and the differential modulus is: 
\begin{equation}
\begin{aligned}
K(\omega;\sigma)\equiv \frac{\diff \sigma}{\diff \gamma}=\frac{\rho}{\chi(\omega;F=\frac{\sigma}{\rho})}.
\end{aligned}
\label{e17}
\end{equation}

We start by calculating $K$ in the low-frequency regime, which is governed by the transient crosslinkers, \\$\omega\ll\omega_c(\sigma)$, where $\omega_c(\sigma)$ is the characteristic frequency separating the transient regime and the permanent regime (see Fig.~\ref{Fig.rgm}). The mathematical definition of $\omega_c(\sigma)$ will be given later in this section. In this regime the contribution from long-wavelength modes dominates, allowing us to only use the long-wavelength part of $M_{qq}$, i.e.,~$M_{qq}=a_2q^{2}$. Substituting this into Eq.~(\ref{e18}), we have: 
\begin{align}
\chi(\omega;F) &=\frac{2k_B T(2a_2)^{ 1 /2}}{\pi\kappa^{ 3/ 2}\omega^{ 1/ 2}} \notag\\&\times \bigintssss_0^{\infty} \frac{r^4 \diff r} {(r^2+\frac{F}{F_c(\omega)})\left[r^6+\frac{F}{F_c(\omega)}r^4-i\right] },\label{e19}
\end{align}
where $r = (2\kappa a_2/\omega)^{ 1/ 6}q$ and $F_c(\omega)= (\kappa^2\omega/2a_2)^{1/3}$ is a frequency-dependent characteristic tension. Using the expression in Eq.~(\ref{e19}), the differential modulus is calculated using Eq.~(\ref{e17}). We separate the dynamic modulus into the storage modulus, $K'$, and the loss modulus $K''$, using $K=K'+ i K''$. The dynamic modulus can be expressed in a simple form:

\begin{align}
&K'(\omega;\sigma) = K_c(\omega){\rm Re}\left[g\Big({\sigma}/{\sigma_c(\omega)}\Big)\right]\notag\\
&K''(\omega;\sigma) = K_c(\omega){\rm Im}\left[g\Big({\sigma}/{\sigma_c(\omega)}\Big)\right],\label{e20}
\end{align}

where 

\begin{equation}
\begin{aligned}
\left[g(x)\right]^{-1}= {\bigintssss \dfrac{(q^6+ {x}q^4)\diff q}{(q^2+ {x})^2[(q^6+xq^4)-i]} },
\end{aligned}
\label{e21}
\end{equation}
is a function describing the stress-stiffening behavior of semiflexible polymers, and  
\begin{equation}
\begin{aligned}
\sigma_{c}(\omega)= \gamma_1 \omega^{\frac 1 3}\quad;\quad
K_c(\omega) = \gamma_2 \omega^{\frac 1 2},
\label{e22}
\end{aligned}
\end{equation}
are the characteristic stress and modulus, with $\gamma_1=\rho(\kappa^2/2a_2)^{1/3}$ and $\gamma_2 =\pi \rho\kappa^{ 3/ 2}/(2^{ 3/ 2}a_2^{ 1/ 2}k_BT)$. These characteristic stress and  modulus agree with what has been observed in reconstituted actin networks crosslinked by $\alpha$-actinin-4, and was explained using a phenomenological theory~\cite{Mulla2019}. By rescaling the experimental data using $K_c$ and $\sigma_c$, we collapse the  data of the storage modulus onto a single curve, which is fitted with our theoretical prediction of Eq.~(\ref{e20}), see  Fig.~\ref{Fig.2}. Here we fit the data using only two fitting parameters, $\gamma_1$ and $\gamma_2$, for all curves~(i.e. after fitting one line we predict all the rest)~\footnote{The collapse of the experimental data was previously done in Ref.~{\cite{Mulla2019}} using a phenomenological theory.}. 

We continue with calculating the differential modulus in the entire frequency regime. To do so, we substitute  Eq.~(\ref{e34}) into Eq.~(\ref{e18}) and Eq.~(\ref{e17}), leading to the differential modulus, which is the central result of this paper:
\begin{equation}
\begin{aligned}
&K(\omega;\sigma)=[\tilde\chi_t(\omega;\sigma)+\tilde\chi_p(\omega;\sigma)]^{-1},
\end{aligned}
\label{e37}
\end{equation}
where $\tilde \chi_t$ and $\tilde \chi_p$ are contributions from the transient and permanent modes, respectively. The two terms  are
\begin{align}
\tilde\chi_t(\omega,\sigma)&= D_t\int_0^{1-\Delta r}\diff r \Bigg[\frac{r^4[1-r^3]^{-2}}{\omega_{t}r^2+\omega_{t\sigma}}\notag\\&\times\frac{ 1 }  {r^4[1-r^3]^{-2}({\omega_t r^2+\omega_{t\sigma} }) 
	-i\omega}\Bigg]\notag\\
\tilde \chi_p(\omega,\sigma)&=D_p\int_{1-\Delta r}^{\infty}\diff r \Bigg[ \frac{r^2 }  {\omega_{p} r^2+\omega_{p\sigma}}\notag\\&\times\frac{1 }  {\omega_{p} r^2+\omega_{p\sigma} -i\omega}\Bigg],\label{e36}
\end{align}
where $r = q/q_c$, $\Delta r = \Delta q /q_c$, $D_t =0.11\gamma_2^{-1}\tau_{\rm off}^{-3/2}$, $\omega_t=0.23/\tau_{\rm off}$, $\omega_{t\sigma}=0.38\gamma_1^{-1}\tau_{\rm off}^{-2/3}\sigma$, $D_p=2.09 \gamma_2^{-1}\tau_{\rm off}^{1/2}\tau_{\rm per}^{-2}$, $\omega_p=1/\tau_{\rm per}$ and $\omega_{p\sigma}=1.63 \gamma_1^{-1}\tau_{\rm per}^{-1}\tau_{\rm off}^{1/3}\sigma$.  Here $\Delta q = {{a_2}^{ 1/ 2}q_c^2}/({3M_0^{1/ 2}})$ as defined after Eq.~(\ref{e32}). The nonlinear modulus for any given $\omega$ and $\sigma$ is then determined by four parameters: $\gamma_1$, $\gamma_2$, $\tau_{\rm off}$ and $\tau_{\rm per}$. 

The transient and permanent regimes are determined by the dominant term in  Eq.~(\ref{e37}) (when $\tilde\chi_t$ dominates the network is in the transient regime and when $\tilde\chi_p$ dominates it is in the permanent regime). We then define the characteristic frequency $\omega_c$ to be the one satisfying $|\tilde \chi_t(\omega_c,\sigma)|=|\tilde\chi_p(\omega_c,\sigma)|$. For small stress, numerical analysis suggests $\omega_c(\sigma=0)=1.26\omega_t=0.29/\tau_{\rm off}$, indicating that the unbinding time is the timescale separating the transient regime \SC{ from the permanent regime}. For larger stress $\omega_c$ is decreasing and it vanishes as the stress exceeds a threshold (see Fig.~\ref{Fig.rgm}). This suggests that for very large stress the transient behavior vanishes, and the permanent-regime plateau expands and covers the entire low-frequency regime. However, such a shift is only obvious when $\sigma>\sigma_p$, where $\sigma_p$ is the characteristic stress for the permanent regime, and $\omega_c$ can be regarded as a constant in the linear regime (see Fig.~\ref{Fig.rgm}). 

In the transient regime, i.e. $\omega\ll\omega_c(\sigma)$, $\tilde\chi_t$  dominates Eq.~(\ref{e37}), and the dynamic moduli are well described by Eq.~(\ref{e20}). For small prestress we have $K\sim \omega^{1/2}$, same as the linear modulus. When the prestress increases the term $r^4(\omega_t r^2+\omega_{t\sigma})$ is comparable to $\omega$, and the network  stiffens nonlinearly. The characteristic stress for this stiffening is $\sigma_c$, see Eq.~(\ref{e22}). Since $\sigma_c\sim \omega^{1/3}$, the characteristic stress is larger for higher frequency, and the frequency dependence of $K$ is also weakened by the prestress. In fact, the apparent scaling exponent of $K$ can be written as a function of $\sigma/\sigma_c$, see Sec.~\ref{section:6} for further discussion. 

When $\omega\gg\omega_c(\sigma)$, $\tilde\chi_p$  dominates  Eq.~(\ref{e36})  and the network is in its permanent regime. In this case the network behaves as the well-studied permanent network~\cite{Gittes1998,Morse1998}. There is a noteworthy frequency in Eq.~(\ref{e37}) associated with the permanent regime, $\omega_p+\omega_{p\sigma}$, which is the relaxation rate for the bending modes with $q$ slightly larger than $q_c$. As we assume $\tau_{\rm off}\gg \tau_{\rm per}$, we have $\omega_c(\sigma)\ll\omega_p+\omega_{p\sigma}$, and we expect a plateau in $K'$ between the two frequencies (see Ref.~\cite{Gittes1998} and Appendix B). In fact, when $\omega_c(\sigma)\ll\omega\ll\omega_p+\omega_{p\sigma}$ in Eq.~(\ref{e37}), we have $K\approx C_1 + (C_2/\omega)i$, where $C_1$ and $C_2$ do not depend on $\omega$.  The different scaling for $K'$ and $K''$ in this plateau regime results in different behavior in the transition between the transient and permanent regimes, see discussion on Fig.~\ref{Fig.3} below. For $\omega\gg\omega_p+\omega_{p\sigma}$, we have $K\sim \omega^{3/4}$, in agreement with analytical results for permanent networks~\cite{Gittes1998, Morse1998}. The network in the permanent regime nonlinearly stiffens when the prestress reaches the characteristic stress, $\sigma_p$, which is defined as the prestress satisfying $\omega_p=\omega_{p\sigma}$.

 We then determine the four parameters ($\gamma_1, \gamma_2, \tau_{\rm off}, \tau_{\rm per} $) by fitting the experiment data of Ref.~\cite{Mulla2019}. We find $\gamma_1$ and $\gamma_2$ by fitting the $K'$ data with Eq.~(\ref{e20}) (Fig.~\ref{Fig.2} (b)), then fix these values and fit the $K''$ data to find $\tau_{\rm off}$ and $\tau_{\rm per}$ (Fig.~\ref{Fig.3} (b)), where we minimize the sum of the squared deviation of the theoretical prediction (Eq.~\ref{e37}) from the experimental data for all curves simultaneously (each curve with different $\sigma$). The same values of $\gamma_1$, $\gamma_2$, $\tau_{\rm off}$ and $\tau_{\rm per}$ are used for the entire family of curves. Using the best-fit values of the parameters, we find that $\tau_{\rm off} =0.0085\,{\rm s}$, in agreement with previous result~\cite{Mulla2019}. The value of $\tau_{\rm per}$ may be inaccurate, since the experimental data is only in the transient and intermediate regimes. Changing the value of $\tau_{\rm per}$ only slightly affects the predicted moduli, as long as $\tau_{\rm per}\ll \tau_{\rm off}$. 

As shown in Fig.~\ref{Fig.3}(a), the fitting for the $K'$ data is excellent. Our theory predicts that increasing the prestress results in an increase of $K'$ together with a weakening of its frequency dependence, as is also observed experimentally. For $K''$ the theory does not agree well with the experimental data. Although our theory shows the same qualitative features as the experiments, including the strengthened loss modulus and decreased scaling exponent for $K''(\omega)$, our predicted  $K''$ is always smaller than experimentally observed in the low-frequency regime (see Fig.~\ref{Fig.3}(b)). We believe that one reason for this is the faster increase of $K'$ with prestress (compared to $K''$). Then, for high prestress, the ratio $K''/K'$ can be close to $0.1$, making it experimentally hard to get accurate  measurement of $K''$ (see inset of Fig.~\ref{Fig.3}(b)).  This is a known issue that was discussed in Ref.~\cite{Velankar2007}. The other reason for the disagreement is that the mobility we used for $q\sim q_c$ can be inaccurate: the mobility for $q\gg q_c$ and $q\ll q_c$ is well understood, but the transition between the two parts  was not explored in this work. The prediction for $K''$ is more affected by this transition because the two regimes are well separated for $K''$: $K''$ increases with frequency in the transient regime but decreases $\sim\omega^{-1}$  in the permanent regime.  On the contrary, $K'$ undergoes an insignificant change because after increasing with frequency in the transient regime it reaches a plateau in the permanent regime, and is thus less affected by the transition itself. 
\begin{figure}[t]
	\centering
	\includegraphics[scale=0.93]{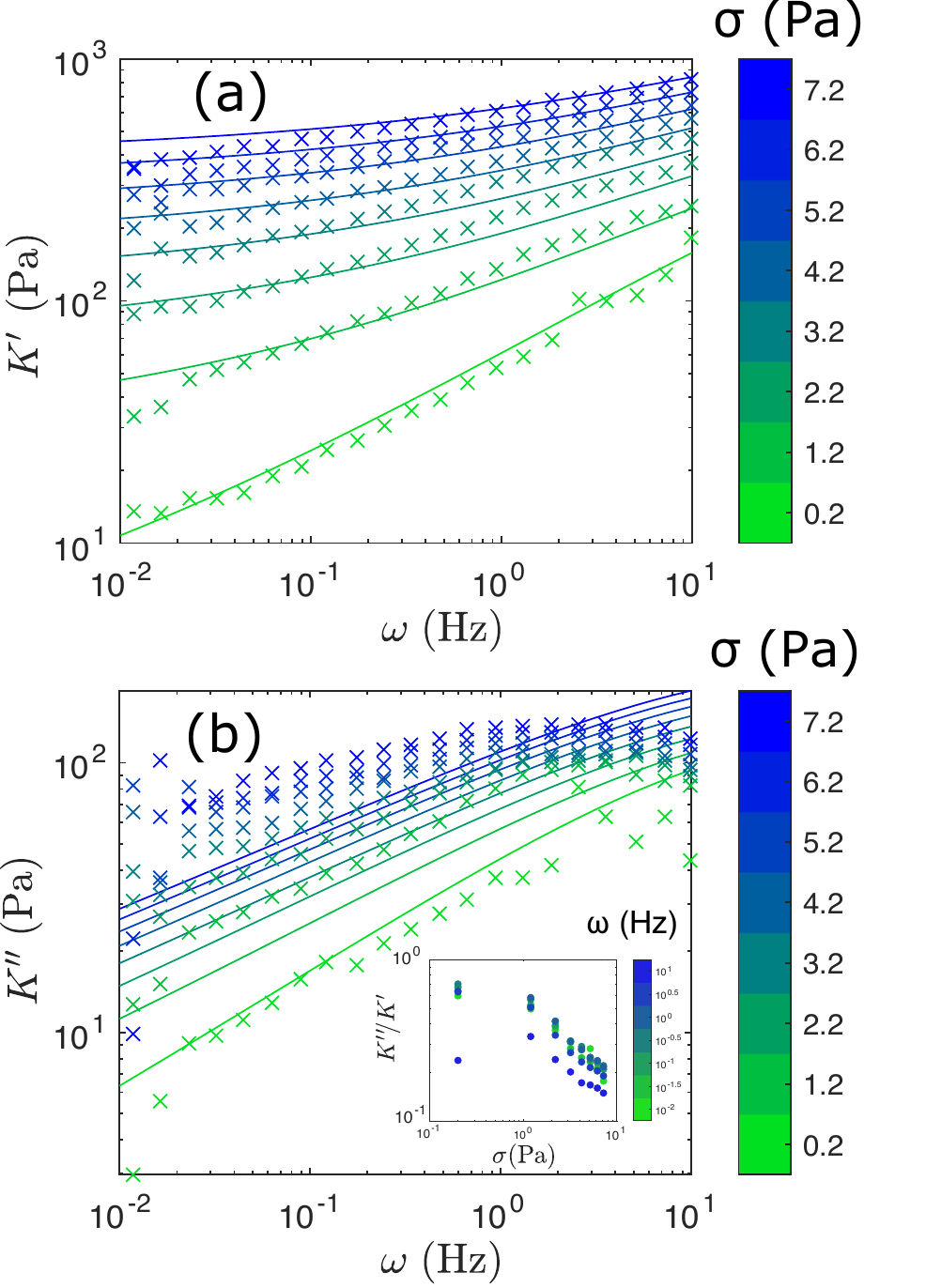}
	\caption{Differential storage modulus $K'$ (a) and differential loss modulus $K''$ (b) as measured in the experiments of Ref.~\cite{Mulla2019} (x symbols), together with the  theoretical fitted curves of Eq.~(\ref{e37}). The best-fit values of the parameters are: $\gamma_1 = 1.71 \,{\rm Pa \, s^{\ 1/ 3}}$, $\gamma_2 = 35.9\,{\rm Pa \, s^{ 1/ 2}}$, $\tau_{\rm off}=0.0085\, {\rm s}$ and $\tau_{\rm per} = 2.0\times 10^{-14} \,{\rm s\,}$. In the inset of (b) we plot the ratio $K''/K'$ measured experimentally as function of $\sigma$. }
	\label{Fig.3}
\end{figure}

\section{Discussion and Conclusion}
\label{section:6}

We have proposed a general theory for the nonlinear modulus of biopolymer networks, which describes the dynamics of the networks using a mode-dependent mobility, $M_{qq}$. In transient-crosslinked networks, we find that $M_{qq}\sim q^2$ for the long-wavelength modes, indicating that the relaxation of  these modes is slowed down by transient crosslinkers. This explains both the $1/2$ scaling exponent of the frequency-dependent linear modulus reported in Ref.~\cite{Broedersz2010}, and the glassy-like relaxation in the presence of prestress~\cite{Mulla2019}. Our theory suggests that the transient nature of crosslinkers in  biopolymer networks is the cause of the apparent weak exponent of the frequency-dependent modulus found in living cells~\cite{Fabry2001}.

\begin{figure}[t]
	\centering
	\includegraphics[scale=0.6]{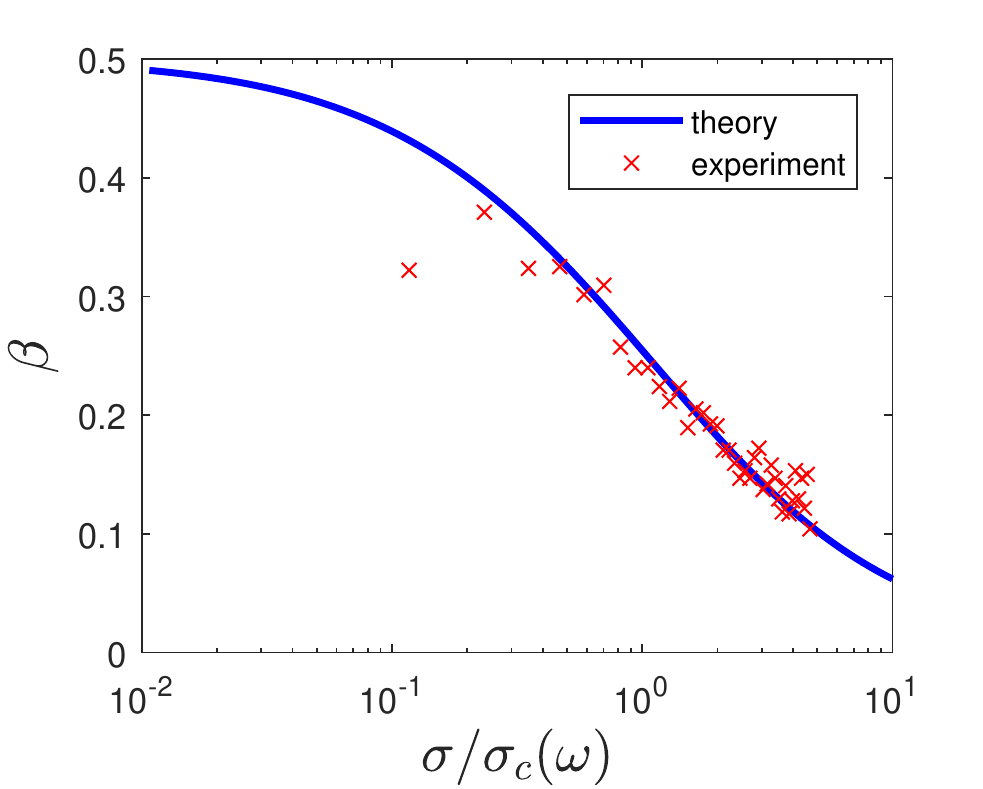}
	\caption{Apparent scaling exponent $\beta$ as function of stress measured in experiment, together with our theoretical prediction of Eq.~(\ref{e23}). Experimental data is taken from Ref.~\cite{Mulla2019}, with $\omega=0.31 {\rm Hz}$.}
	\label{Fig.5}
\end{figure}

To further analyse how the prestress affects the frequency dependence of the nonlinear modulus, we calculate the apparent scaling exponent $\beta(\omega,\sigma)$ around a given frequency $\omega$ when the network is subjected to a prestress $\sigma$:

\begin{equation}
\begin{aligned}
\beta(\omega;\sigma)=\frac{\partial \ln K'(\omega;\sigma)}{\partial \ln \omega}\,.
\end{aligned}
\label{e23}
\end{equation}
In the transient regime where Eq.~(\ref{e20}) is valid, $\beta$ is only a function of $\sigma/\sigma_c(\omega)$.  In Fig.~\ref{Fig.5} we plot   $\beta$ for varying $\sigma/\sigma_c(\omega)$ and fixed $\omega=0.31\,{\rm Hz}$, and find that the apparent scaling exponent $\beta$ decreases with increasing prestress. Further, $\beta$ can be  tuned to any value between $0$ and $1/2$, depending on the prestress strength. This might explain the weak scaling exponent found in living cells. The cytoskeleton, which gives the cell its rigidity, is formed by many transient crosslinking proteins, and is effectively a prestressed transient biopolymer network, where molecular motors are responsible for the prestress \cite{Schiffhauer20161473}. Moreover, if the crosslinkers binding/unbinding process is out of  equilibrium, an internal stress can be created even in the absence of molecular motors~\cite{Chen2020}. Experiments on living cells have also observed reduced scaling exponent for increasing internal motor stress~\cite{Stamenovic2004,Kollmannsberger20113127}.

The weak scaling exponent of the frequency-dependent shear modulus in living cells has been discussed repeatedly during the past two decades ~\cite{Fabry2001,Stamenovic2004,Desprat20052224,Bursac2005557,Balland2006}. Most of the previous works try to explain this weak exponent using the soft glassy rheology (SGR) model, which describes the dynamics of soft materials that have structural disorder and metastability ~\cite{Fabry2001,Bursac2005557}. These materials are out of equilibrium, as thermal energy is insufficient to drive the systems across the energy barrier to reach equilibrium~\cite{Navarro1996, Sollich1998,Cloitre2003}. Although living cells are also out of equilibrium,  there is no direct evidence that the origin of the reported weak exponent in living cells is the same as that of soft glassy materials. \SC{Also, the scaling exponent in the SGR model does not change with prestress, which does not agree with most experiments on living cells as well as the more recent experiments on reconstituted networks ~\cite{Stamenovic2004,Kollmannsberger20113127,Mulla2019}. }

In this paper, we describe the dynamics of transient-crosslinked biopolymer networks and provide the microscopic understanding of the nonlinear transient regime. Although our work treats a system close to thermal equilibrium, we find that it exhibits the same weak-scaling phenomena. Here the weak exponent is a result of the coupling between multiple relaxation times, which comes from the relaxation of infinite bending modes that are slowed down by the transient crosslinkers. Our theory suggests that a distribution of long relaxation times can exist in an equilibrium system with only short obvious timescales, where the long relaxation times come from the collective behavior of multiple crosslinks.  Unlike glassy systems with metastability, our system does have a single longest relaxation time. This relaxation time is not the system intrinsic time-scale $\tau_{\rm off}$, which is related to the binding/unbinding process, rather it is the relaxation time for the longest-wavelength mode, $q = \pi/\ell$ (Eq.~\ref{e6}): $\tau_r \sim \tau_{\rm off}\ell^6/\ell_c^6$. When the polymers have sufficiently long contour length $\ell$, $\tau_r$ can become very long such that the system is in a nearly glassy state.

In this work we assumed that the mode-dependent mobility and $a_2$ for transient networks are independent of the external stress. This simple assumption is sufficient to explain the experimental data of Ref.~\cite{Mulla2019}, although in general, $a_2$ may be a function of the polymer tension. For example, the lifetimes of the bound states may depend on the mechanical force exerted on the crosslinkers, which can lead to a change in $a_2$. For most protein complexes this unbinding time decreases under external force, as they are effectively dragged off. Such an effect is called slip bond. The opposite behavior, catch bond, also exists in some crosslinking proteins~\cite{Schiffhauer20161473}, including $\alpha$-actinin-4 that is used in Ref.~\cite{Mulla2019}. To account for this change in the unbinding time, we can replace the constant $a_2$ with $a_2(\sigma)$, where the rest of the derivation remains unaffected. For slip bond, we expect $a_2$ to increase with $\sigma$ and the exponent to be smaller than in Fig.~\ref{Fig.5}, while for catch bond we expect the opposite.

Our results are focused on transient networks, but the mode-dependent mobility theory we present is applicable for  other biopolymer networks, such as entangled networks or networks with multiple types of crosslinkers.  Different networks correspond to different mode-dependent mobility, $M_{qq}$. Our theory suggests a one-to-one relation between $M_{qq}$ and the correlation function of a single filament. Since  the linear modulus of the network is usually related to the single-filament correlation function~\cite{Gittes1998}, our theory also implies an underlying relation between the linear and the nonlinear modulus in biopolymer networks.

\begin{acknowledgements}
	\emph{Acknowledgements}:
	This work was supported in part by the National
	Science Foundation Division of Materials Research
	(Grant No.\ DMR-1826623) and the National Science
	Foundation Center for Theoretical Biological Physics
	(Grant No.\ PHY-2019745). The authors
	acknowledge helpful discussions with G.~Koenderink and Y.~Mulla.
\end{acknowledgements}

\appendix
\section{Mean-field Theory of the Coursed-Grained Dynamics (CGD) Model}
\renewcommand{\theequation}{A\arabic{equation}}
\setcounter{equation}{0}
\renewcommand{\thefigure}{A\arabic{figure}}
\setcounter{figure}{0}
In this appendix we present the complete derivation of the mean-field theory proposed in Ref.~\cite{Broedersz2010}, with the goal of understanding the dynamics of the end-to-end distance of a single polymer, and deriving the end-to-end response function $\chi(\omega)$, in a transient-crosslinked network. 

To analytically solve the end-to-end dynamics, we treat a single polymer in a transient network using a coarse-grained dynamics model (CGD)~\cite{Broedersz2010}. Within this model, a crosslinked polymer of length $\ell$ with average crosslinking distance $\ell_c$, is treated as $N=\ell/\ell_c$ polymer segments of length $\ell_c$ separated by crosslinkers on a 2D plain, see Fig.~\ref{Fig.4}. Each segment is modeled as an entropic spring with stretching rigidity $\mu_{\rm th}=180\kappa^2/(k_B T \ell_c^3)$~\cite{Broedersz2014} and the bending interactions between adjacent segments is considered via the bending rigidity $\kappa$.  Overall, the Hamiltonian of the entire chain can be written as:
\begin{equation}
\begin{aligned}
H_{\rm CG} = \frac 1 {\ell_c}\sum_{n=1}^N\left[\frac {\mu_{th}} 2 (|\Delta {\bf r}_n|-\ell_c )^2+\frac \kappa 2|\theta_n|^2\right]\,,
\end{aligned}
\label{S1}
\end{equation}
where $\Delta {\bf r}_n={\bf r}_n-{\bf r}_{n-1}$, ${\bf r}_n$ being the position of the $n^{\rm th}$ crosslinker and $\theta_n$ is the angle between the $n^{\rm th}$ segment and the $n+1$ segment (see Fig.~\ref{Fig.4}). In the semiflexible limit ($\kappa/k_B T\gg \ell_c$) where $\theta_n$ is small, we have $\theta_n=|{\bf \hat t}_n-{\bf \hat t}_{n-1}|$ with ${\bf \hat t}_n=\Delta {\bf  r}_n/|\Delta {\bf  r}_n| $.

\begin{figure}[t]
	\centering
	\includegraphics[scale=0.38]{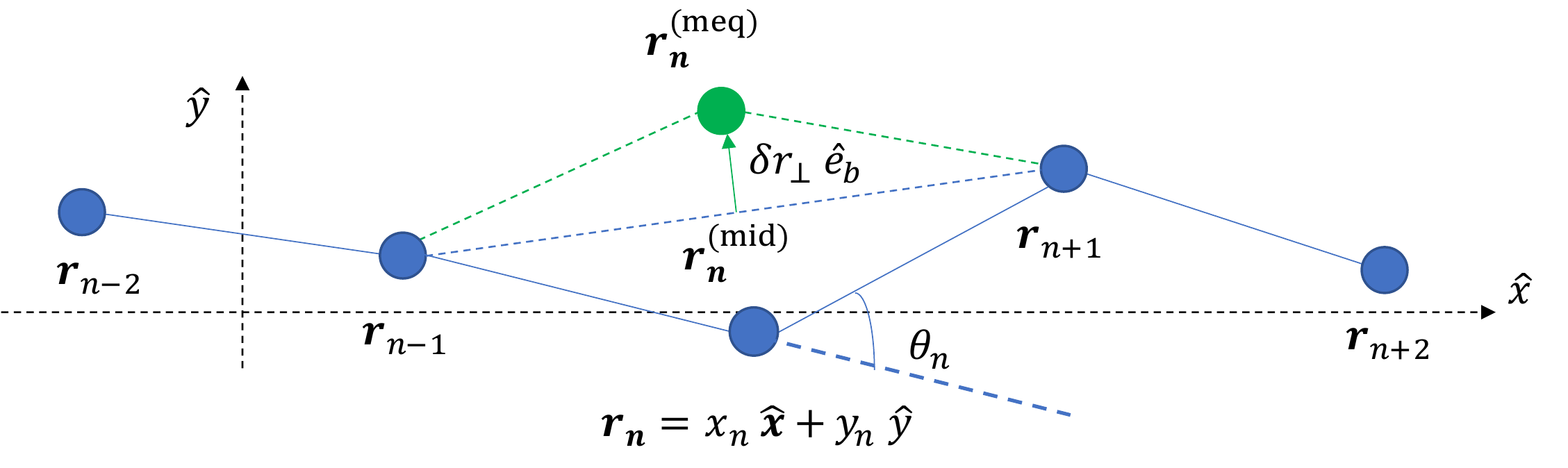}
	\caption{Illustration of the CGD model. A semiflexible polyer is moving on a 2D plain, where $\hat{x}$  is the longitudinal direction and $\hat{y}$ is the transverse direction.  Each blue node represents a crosslinker and each solid line represents a polymer segment. The green node is the mechanical equilibrium position of node $n$.    }
	\label{Fig.4}
\end{figure}

We use 2D Cartesian coordinate to describe the position of each crosslinker in the polymer, where ${\hat x}$  is the direction of the polymer backbone and ${\hat y}$  denotes the transverse direction~(see Fig.~\ref{Fig.4}), such that ${\bf r}_n=x_n \, \hat x+y_n \, \hat y$. Since the transverse fluctuations of semiflexible polymers are small, we have $|y_n-y_{n-1}|\ll\ell_c$. Because the semiflexible polymers are hard to stretch, the length scale of the end-to-end fluctuation is also much smaller than $\ell_c$, $\Big||\Delta {\bf r}_n|-\ell_c\Big|\ll\ell_c$, hence, ${\bf \hat t}_n\simeq [1,(y_n-y_{n-1})/\ell_c]$. Using the above conditions, we can approximate $\Delta {\bf r}_n$ and $\theta_n$ as:
\begin{equation}
\begin{aligned}
|\Delta {\bf r}_n|&=\sqrt{(x_n-x_{n-1})^2+(y_n-y_{n-1})^2} \\&\simeq(x_n-x_{n-1})+\frac{(y_n-y_{n-1})^2}{2\ell_c}\,,\\
|\theta_n|&\simeq\frac 1 \ell_c |y_{n+1}-2y_n+y_{n-1}|\,.
\end{aligned}
\label{S2}
\end{equation}

Having presented the Hamiltonian of the CGD model, let us consider the polymer stress relaxation process. The polymer is connected to the rest of the network via the transient crosslinkers, i.e. the nodes in our model. We assume these crosslinkers have strong binding affinity such that $\tau_{\rm on}\ll\tau_{\rm off}$, where $\tau_{\rm on}$ and $\tau_{\rm off}$ are the average lifetimes in the unbound and bound states, respectively.  Then, it is unlikely that two nodes will be unbound at the same time. Furthermore, since the CGD model treats networks of semiflexible polymers, it is appropriate to assume that a crosslinker in the bound state cannot change its position (i.e. it is connected to a rigid network). In the unbound state the crosslinker is free to move and relax the stress. Assuming that $\tau_{\rm eq}\ll\tau_{\rm on}$, where $\tau_{\rm eq}$ is the relaxation time of the node, after a node unbinds, it relaxes completely according to $H_{\rm CG}$ before it binds again, while all other nodes positions remain unchanged. The entire chain can then deform through successive unbinding/rebinding events. 

We continue by considering a single unbinding/rebinding event of the $n^{\rm th}$ crosslinker, which changes its position   from $\bf{r}_n$ (before unbinding) to $\bf{r}_n^*$ (after rebinding), while the positions of all other nodes, $\{{\bf r}_{m \neq n}\}$, remain fixed.  We expect ${\bf r}_n^*$ to fluctuate around its mechanical equilibrium position, ${\bf r}_n^{\rm meq}$, which is determined from the  force balance equation:
\begin{equation}
\begin{aligned}
\frac{\partial H_{\rm CG}}{\partial {\bf r}_n}\Bigg|_{{\bf r}_n={\bf r}_n^{\rm meq}}=0\,.
\end{aligned}
\label{S37}
\end{equation}
 We first consider the $\hat x$ component of Eq.~(\ref{S37}):
\begin{align}
\frac{\partial H_{\rm CG}}{\partial x_n}\Bigg|_{{\bf r}_n={\bf r}_n^{\rm meq}}&=& \frac {\mu_{th}} {\ell_c}\Bigg[\left|{\bf r}_n^{\rm meq}-{\bf r}_{n-1}\right|^2\notag\\&\,\,-&|{\bf r}_n^{\rm meq}-{\bf r}_{n+1}|^2\Bigg]=0\,.\label{S3}
\end{align}
Equation (\ref{S3}) suggests that ${\bf r}_n^{\rm meq}$ has the same distance to ${\bf r}_{n-1}$ and ${\bf r}_{n+1}$, and is therefore located at the angular bisector of ${\bf r}_{n-1}$ and ${\bf r}_{n+1}$. This means we can write ${\bf r}_n^{\rm meq}$ as ${\bf r}_n^{\rm meq}={\bf r}_n^{\rm mid} + \delta {\bf r}$, where ${\bf r}_n^{\rm mid}$ is the midpoint of ${\bf r}_{n-1}$ and ${\bf r}_{n+1}$ and $\delta{\bf r}=\delta x \, \hat x+\delta y \hat y=|\delta r| {\bf\hat e}_b$ is perpendicular to $({\bf r}_{n+1}-{\bf r}_{n-1})$. We next consider the $\hat y$ component of Eq.~(\ref{S37}):
\begin{align}
&\frac{\partial H_{\rm CG}}{\partial y_n}\Bigg|_{{\bf r}_n={\bf r}_n^{\rm meq}}\!\!\!\!\!\!= \frac {\mu_{th}} {\ell_c^2}\left[\delta y \left( d_n-2\ell_c+\frac{|\delta {\bf r}|^2}{\ell_c}\right)\right]\notag\\&\quad+\frac{\kappa}{\ell_c^3}[y_{n-2}-y_{n-1}+6\delta y-y_{n+1}+y_{n+2}]=0\,,\label{S4}
\end{align}
where $d_n =x_{n+1}-x_{n-1}+(y_{n+1}-y_{n-1})^2/8\ell_c $ is the distance between ${\bf r}_{n-1}$ and ${\bf r}_{n+1}$. Equation (\ref{S4}) is composed of a stretching term proportional to $\mu_{\rm th}$ and a bending term proportional to $\kappa$. For a semiflexible polymer,  $\mu_{\rm th}= 180\kappa^2/k_B T \ell_c^3\gg \kappa/\ell_c^2$, hence, the stretching term dominates, and the bending term can be neglected such that
\begin{equation}
\begin{aligned}
|\delta {\bf r} | = \sqrt 2 \ell_c\sqrt{1-\frac{d_n}{2\ell_c}}\Theta\left(1-\frac{d_n}{2\ell_c}\right)\,,
\end{aligned}
\label{S5}
\end{equation}
where $\Theta(x)$ is the heaviside function. 

In the presence of thermal fluctuations the position of the $n^{\rm th}$ crosslinker  ${\bf r}_n^*$ will fluctuate around its mechanical equilibrium position ${\bf r}_n^{\rm meq}$. We define this deviation  to be ${\bm \xi}= {\bf r}_n^*-{\bf r}_n^{\rm meq}$. Because $\mu_{\rm th}$ is large, the deviation is approximately parallel to  ${\bf \hat e}_b$, namely,  ${\bm \xi}\times({\bf r}_{n+1}-{\bf r}_{n-1})\simeq 0$, and ${\bm \xi}=\xi_b \,{\bf \hat e}_b$. Assuming the deviation from mechanical equilibrium is small, we expand the Hamiltonian of Eq.~(\ref{S1}) to  second order in $\xi_b$:
\begin{align}
H_{\rm CG}({\bf r}_n^*,\{{\bf r}_{m\neq n}\})\!-\!\!H_{\rm CG}({\bf r}_n^{\rm meq},\{{\bf r}_{m \neq n}\})&= \frac{\mu_{\rm th}}{\ell_c^3}|\delta {\bf r}|^2\xi_b^2\notag\\&+O(\xi_b^3)\,.\label{S6}
\end{align}
Since node $n$ completely relaxes before rebinding to the substrate, ${\bf r}_n^*$ follows a Boltzmann distribution governed by $H_{\rm CG}({\bf r}_n^*,\{{\bf r}_{m\neq n}\})$, while the positions of other nodes remain fixed. We denote this distribution as $P_n({\bf r}_n^*|\{{\bf r}_{m \neq n}\})$, and  the average with respect to  $P_n({\bf r}_n^*|\{{\bf r}_{ m \neq n}\})$ is denoted by $\langle ...\rangle_{n}$. According to Eq.~(\ref{S6}), we have $\langle \xi_b\rangle_n=0$ and $\langle \xi_b^2\rangle_n=k_BT\ell_c^3/(2\mu_{\rm th}|\delta {\bf r}|^2)$, such that $\xi_b$ can be considered as a thermal noise. 

Our goal is to calculate the polymer's end-to-end response function. To that aim, it is sufficient to calculate the parallel component of a node displacement in one unbinding/rebinding event. The parallel displacement of a node follows
\begin{equation}
\begin{aligned}
x_n^* = \frac 1 2 (x_{n-1}+x_{n+1})+ \hat x \cdot (\delta {\bf r}+ \xi_b \hat{\bf e}_b)\,,
\end{aligned}
\label{S7}
\end{equation}
where ${\bf r}_n^*=x_n^*\,\hat{x} + y_n^*\,\hat{y}$. As both $\delta {\bf r}$ and $\xi_b$ depend explicitly on both the $\hat x$ and $\hat y$ components, Eq.~(\ref{S7}) cannot be solved without considering the perpendicular component of the node displacement. To simplify the equation, we use a mean-field approach which replaces the terms containing $\hat{\bf e}_x \cdot \delta {\bf r}$ and $\xi_b\hat{\bf e}_x \cdot \hat{\bf e}_b$ with their long-time average values, namely an average over many unbinding/rebinding events. Since the polymer is in thermal equilibrium, this long-time average is equivalent to the average with respect to a Boltzmann distribution of all node positions governed by Eq.~(\ref{S1}), $P_{\rm eq}(\{{\bf r}_{i}\})$. We denode this average by  $\langle ...\rangle _{\rm MF}$. Note that since $P_n({\bf r}_n^*|\{{\bf r}_{ m \neq n}\})$ is also a Boltzmann distribution of ${\bf r}_n^{*}$ with fixed $\{{\bf r}_{m \neq n}\}$, we have $P_{\rm eq}(\{{\bf r}_{i}\})=P_n({\bf r}_n^*|\{{\bf r}_{m \neq n}\})P_{m \neq n}(\{{\bf r}_{m \neq n}\})$, where $P_{m \neq n}(\{{\bf r}_{m \neq n}\})=\int \diff {\bf r}_n P_{\rm eq}(\{{\bf r}_{i}\})$ is the marginal Boltzmann distribution of $\{{\bf r}_{m \neq n}\}$.  Therefore, we have 
\begin{equation}
\begin{aligned}
\langle ...\rangle_{\rm MF}=\langle \langle...\rangle_{n}\rangle_{m \neq n}\,,
\end{aligned}
\label{S38}
\end{equation}
where $\langle ...\rangle_{m \neq n}$ is the average with respect to $P_{m \neq n}$. Since $P_{m \neq n}$ is the marginal distribution of $P_{\rm eq}$, for any variable $A$ that is not a function of ${\bf r}_n^*$, we have $\langle A\rangle_{m \neq n}=\langle A\rangle_{\rm MF}$.

Let us start with averaging the $\hat{\bf e}_x \cdot \delta {\bf r}$ term. Since  ${\bf\hat e}_b$ is  perpendicular to $({\bf r}_{n+1}-{\bf r}_{n-1})$,  ${\bf\hat e}_b$ can point in only two directions. Due to the polymer symmetry in these two directions,  we have $\langle\hat{\bf e}_x \cdot \delta {\bf r} \rangle_{\rm MF}=0$.

Next, we calculate the mean-field average \\$\langle(\xi_b\hat{\bf e}_x \cdot \hat{\bf e}_b)^2 \rangle_{\rm MF}$ using Eq.~(\ref{S38}), \\$\langle(\xi_b\hat{\bf e}_x \cdot \hat{\bf e}_b)^2 \rangle_{\rm MF}=\langle\langle(\xi_b\hat{\bf e}_x \cdot \hat{\bf e}_b)^2 \rangle_{\rm n}\rangle_{m \neq n}$. Because \\$|\hat{\bf e}_x \cdot  \hat{{\bf e}_b}|=|(y_{n+1}-y_{n-1})/(2\ell_c)|$ does not depend on ${\bf r}_{n}^*$, we have
\begin{equation}
\begin{aligned}
\langle \xi_{\rm MF}^2\rangle&\equiv\langle\langle\xi_b^2 \rangle_{\rm n}|\hat{\bf e}_x \cdot \hat{\bf e}_b|^2\rangle_{m \neq n}=\langle\langle\xi_b^2 \rangle_{\rm n}|\hat{\bf e}_x \cdot \hat{\bf e}_b|^2\rangle_{\rm MF}\,,
\end{aligned}
\label{S39}
\end{equation}
where $\langle\xi_b^2 \rangle_{\rm n}$ is determined by $d_n$ (see Eq.~(\ref{S5})). For a semiflexible polymer, $d_n\simeq |\Delta {\bf r}_n|+|\Delta {\bf r}_{n+1}|$ and $|\hat{\bf e}_x \cdot \hat{\bf e}_b|^2$ is determined by $|y_{n+1}-y_{n-1}|$, which can be calculated from $|\theta_n|$ using Eq.~(\ref{S2}). Since the Hamiltonian of Eq.~(\ref{S1}) is quadratic, $\{|\Delta {\bf r}_n|\}$ and $\{|\theta_n|\}$ are uncorrelated, and therefore $\langle\xi_b^2 \rangle_{\rm n}$ and $|\hat{\bf e}_x \cdot \hat{\bf e}_b|^2$ are uncorrelated. We can then write
\begin{equation}
\begin{aligned}
\langle \xi_{\rm MF}^2\rangle={\langle\langle\xi_b^2 \rangle_{\rm n}\rangle_{\rm MF}}{\langle|\hat{\bf e}_x \cdot \hat{\bf e}_b|^2\rangle_{\rm MF}}\,.
\end{aligned}
\label{S40}
\end{equation}

The quadratic form of the Hamiltonian (Eq.~(\ref{S1})) also suggests that $(|\Delta{\bf r}_n|-\ell_c)$ is a Gaussian variable with $\langle |\Delta{\bf r}_n|-\ell_c \rangle_{\rm MF}=0$  and $\langle (|\Delta{\bf r}_n|-\ell_c)^2 \rangle_{\rm MF}=k_B T \ell_c/\mu_{\rm th}$, such that $\langle d_n-2\ell_c \rangle_{\rm MF}=0$  and \\$\langle (d_n-\ell_c)^2 \rangle_{\rm MF}=2k_B T \ell_c/\mu_{\rm th}$. According to  Eq.~(\ref{S6}) we have
\begin{equation}
\begin{aligned}
\langle\langle\xi_b^2 \rangle_{\rm n}\rangle_{\rm MF} = \frac{k_B T\ell_c^3}{2\mu_{\rm th}\langle |\delta {\bf r}|^2\rangle_{\rm MF}}\,,
\label{S41}
\end{aligned}
\end{equation} 
where the mean-field value of $|\delta {\bf r}|^2$  is found using Eq.~(\ref{S5}), $\langle |\delta {\bf r}|^2\rangle_{\rm MF}= 0.5\sqrt{k_B T\ell_c^3/\mu_{\rm th}} $, leading to  
\begin{equation}
\begin{aligned}
\langle\langle\xi_b^2 \rangle_{\rm n}\rangle_{\rm MF} =\sqrt{\frac{{k_B T\ell_c^3}}{{\mu_{\rm th}}}}\,.
\label{S21}
\end{aligned}
\end{equation}
The mean-field value $|\hat{\bf e}_x \cdot  {\hat{\bf e}_b}|=|(y_{n+1}-y_{n-1})/2\ell_c|$ is most easily calculated in the continuum limit  in which the transverse  and parallel displacements are the continuum functions $r_{\perp}(s)$ and $r_{\parallel}(s)$, respectively. Here $s$ is the original parallel position of the node, i.e, $s=n\ell_c$ for node $n$, while $r_\perp(s=n \ell_c)=y_{n}$ and $r_\parallel(s=n \ell_c)=x_{n}$. In this continuum limit $|\hat{\bf e}_x \cdot  {\hat{\bf e}_b}|=|\partial_sr_{\perp}(s=n\ell_c)|$, and the bending part of the Hamiltonian of  Eq.~(\ref{S1}) can be written as: 
\begin{equation}
\begin{aligned}
H_{\rm bend}=\frac{\kappa}{2}\int \diff s \,\left(\frac{\partial^2 r_\perp}{\partial s^2}\right)^2\,.
\end{aligned}
\label{S16}
\end{equation}
Equation (\ref{S16}) is diagonalized using the  Fourier series of $r_{\perp}(s)$~\cite{Gittes1998,Morse1998,Broedersz2010} (assuming periodic boundary conditions):
\begin{equation}
\begin{aligned}
\!\!\!\!\!\!\!\!\!r_\perp(s) &= \sum_q[u_q \sin(qs)+v_q\cos(qs)] \quad\left(q=\frac{2n\pi}{\ell}\right),
\end{aligned}
\label{S17}
\end{equation}
where $n=1,2,3...$. All $u_q$'s and $v_q$'s are Gaussian variables satisfying $\langle u_q\rangle_{\rm MF}=\langle v_q \rangle_{\rm MF}=0$ and \\$\langle u_q^2\rangle_{\rm MF}=\langle v_q^2\rangle_{\rm MF}=2k_B T/(\ell\kappa q^4)$. We then have:
\begin{equation}
\begin{aligned}
\langle (\hat{\bf e}_x \cdot  {\hat{\bf e}_b})^2\rangle_{\rm MF} &=\langle (\partial_s r_\perp)^2\rangle_{\rm MF}=\frac{\ell k_BT}{12\kappa}\,.
\end{aligned}
\label{S18}
\end{equation}
Substituting Eq.~(\ref{S21}) and (\ref{S18}) into Eq.~(\ref{S39}) gives
\begin{equation}
\begin{aligned}
\langle \xi_{\rm MF}^2 \rangle = \frac{\ell (k_B T)^{3 /2}\ell_c^{{ 3}/ {2}}}{ 12\kappa \mu_{\rm th}^{ 1/ 2}}\,.
\label{S22}
\end{aligned}
\end{equation}
The mean-field average of Eq.~(\ref{S7}) is then written as:
\begin{equation}
\begin{aligned}
\Delta \bar{x}_n = \frac 1 2 (\bar x_{n-1}-2\bar x_n+\bar x_{n+1}) + \xi_{\rm MF}\,,
\end{aligned}
\label{S8}
\end{equation}
where $\bar{x}_n=\langle x_n\rangle_{\rm MF} $ and $\Delta \bar x_n = \bar x_n^*-\bar x_n$ is the parallel displacement in one unbinding/rebinding event. 

The continuum limit (in both time and space) of Eq.~(\ref{S8}) is then written as
\begin{equation}
\begin{aligned}
\tau_{\rm off}\partial_t \bar r_{\parallel} = \frac{\ell_c^2} 2 \partial^2_s \bar r_{\parallel} + \eta_{\rm MF}\,,
\end{aligned}
\label{S9}
\end{equation}
where the time interval of the unbinding/rebinding event is $\tau_{\rm off}$ and the space interval is $\ell_c$. Here $\bar r_{\parallel}=\langle r_{\parallel}\rangle_{\rm MF}$ and $\langle \eta_{\rm MF}(s,t)\eta_{\rm MF}(s',t')\rangle=\langle\xi_{\rm MF}^2 \rangle\tau_{\rm off}\ell_c \delta(s-s')\delta(t-t')$.  \SC{To solve Eq.~(\ref{S9}), we use the Fourier series: $\bar r_\parallel(s,t) = ({\pi}/{\ell})\sum_q  w(q,t)\exp(iqs)$, where $q=n\pi/\ell$. Following the same reasoning of the main text (see paragraph before Eq.~(\ref{e26})), in the long chain limit we can replace the summation with an integral. The Fourier series is then replaced with $\bar r_\parallel(s,t) = \int \diff q \,w(q,t)\exp(iqs)$, leading to:}
\begin{equation}
\begin{aligned}
\tau_{\rm off}\partial_t w(q,t) = -\frac{\ell_c^2} 2 q^2w(q,t) + \eta(q,t)\,,
\end{aligned}
\label{S10}
\end{equation}
where $\langle\eta(q,t)\eta(q',t')\rangle=\pi\langle\xi_{\rm MF}^2 \rangle\tau_{\rm off}\ell_c\delta(q+q')\delta(t-t')$ and the correlation function is:
\begin{equation}
\begin{aligned}
\langle w(q',t)w(q,0)\rangle = \frac{\pi\langle\xi_{\rm MF}^2 \rangle}{q^2\ell_c}\exp{\left(-\frac{q^2\ell_c^2}{2\tau_{\rm off}}t\right)}\delta(q+q')\,.
\end{aligned}
\label{S11}
\end{equation}

We are interested in the end-to-end extension, \\$\delta \ell = r_\parallel(s=\ell/2)-r_\parallel(s=-\ell/2)-\ell$, which can be found from $w$:
\begin{equation}
\begin{aligned}
\delta \ell(t) = \frac{1}{2\pi}\int \diff q \, w(q,t)(e^{-iq\ell/2}-e^{iq\ell/2})-\ell\,.
\end{aligned}
\label{S12}
\end{equation}
The end-to-end correlation function is then:
\begin{equation}
\begin{aligned}
&\langle \delta\ell(t)\delta\ell(0)\rangle= \frac 1 {\pi}\int \diff q \, \frac{\langle\xi_{\rm MF}^2 \rangle\sin^2(q\ell/2)}{q^2\ell_c}\exp{(-\frac{q^2\ell_c^2}{2\tau_{\rm off}}t)}  \,.
\end{aligned}
\label{S13}
\end{equation}
We now use the  Fourier Transform of Eq.~(\ref{S13}) for the variable $t$ to get the end-to-end power spectrum:
\begin{align}
\langle |\delta\ell(\omega)|^2\rangle& = \frac 1 {\pi}\int \diff q \, \frac{\langle\xi_{\rm MF}^2 \rangle\sin^2(q\ell/2)}{q^2\ell_c}\frac{q^2\ell_c^2/\tau_{\rm off}}{\omega^2+(q^2\ell_c^2/2\tau_{\rm off})^2}\notag\\&\simeq \frac {2 \ell_c \tau_{\rm off}\langle\xi_{\rm MF}^2 \rangle} {\pi}  \, \frac{\diff q}{q^4\ell_c^4+4\tau_{\rm off}^2\omega^2}\,.\label{S14}
\end{align}
In the second step we replaced $\sin^2{(q\ell/2)}$ with its average value, $1/2$. This approximation is valid when $\ell\gg\ell_c/\sqrt{\tau_{\rm off}\omega}$, for which $\sin^2{(q\ell/2)}$ changes with $q$ much faster than the rest of the integrand. 
The linear response function, $\chi(\omega)$, can then be obtained using the fluctuation-dissipation theorem, $\ell \chi''(\omega)=\omega \langle |\delta\ell(\omega)|^2\rangle/2k_B T$, together with the Kramers-Kronig relation:
\begin{align}
\chi(\omega) &= \frac { \ell_c \langle\xi_{\rm MF}^2 \rangle} {2\pi k_B T \ell}\int  \, \frac{\diff q}{q^2\ell_c^2-2i\tau_{\rm off}\omega}\notag\\&=0.0036\frac { k_B T\ell_c^3  } {\pi\kappa^2}\int  \, \frac{\ell_c\diff q}{q^2\ell_c^2-2i\tau_{\rm off}\omega}\,.\label{S15}
\end{align}
Equation (\ref{S15}) is the central result of Ref. \cite{Broedersz2010}, which we use in Eq.~(\ref{e13}) of the main text.  

\section{Discontinuous mode-dependent mobility lead to a plateau}
\renewcommand{\theequation}{B\arabic{equation}}
\setcounter{equation}{0}
\renewcommand{\thefigure}{B\arabic{figure}}
\setcounter{figure}{0}
In the main text we derived the mode-dependent mobility for the transient networks, Eq.~(\ref{e34}), in which a dramatic increase in $M_{qq}$  occurs at $q\sim q_c$. This jump, although seemingly unphysical, is in fact essential for a plateau to appear in the modulus $K'(\omega;\sigma)$, which is observed in experiments of transient networks~\cite{Broedersz2010}.  
In this section we provide a mathematical proof that a plateau in $K'(\omega;\sigma)$ exists if and only if there is a jump (dramatic increase) in the mode-dependent mobility. For simplicity we only discuss the linear modulus, $G'(\omega)=K'(\omega;\sigma=0)$. 

A plateau is defined as a region in which $G'(\omega)$ varies slowly with $\omega$. Since $G(\omega)\sim 1/\chi(\omega)$, a plateau in $G'(\omega)$ also corresponds to a plateau in the real part of the response function $\chi'(\omega)$. Here we calculate $\chi'(\omega)$ using Eq.~(\ref{e18}), 
\begin{equation}
\begin{aligned}
\chi' & =C \int_0^{\infty} \frac{\omega_r(q)^2}{\omega_r(q)^2 + \omega^2 } \frac{d q}{q^4},
\end{aligned}
\label{S32}
\end{equation}
where $C$ is a prefactor and $\omega_r(q)=2\kappa M_{qq}q^4$ is the relaxation rate of mode $q$. To quantify the variation of $\chi'$ for changing $\omega$, we introduce the function, 
\begin{equation}
\begin{aligned}
f(\omega)=\frac{\diff \log(\chi')}{\diff \log(\omega)}=\frac{\omega}{\chi'} \frac{\diff \chi'}{\diff\omega},
\end{aligned}
\label{S33}
\end{equation}
as the local scaling exponent in $\chi'$. The plateau is then identified as a region in which $|f(\omega)|\ll1$.

\begin{figure}[b]
	\centering
	\includegraphics[scale=0.28]{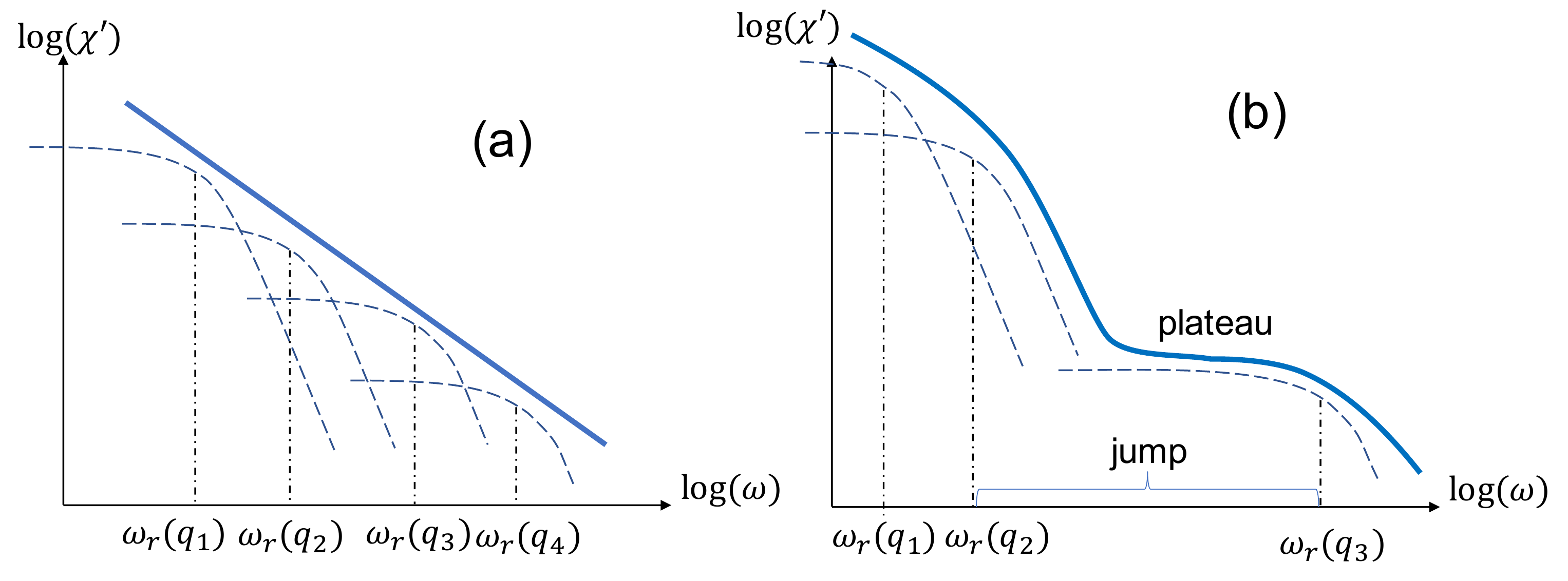}
	\caption{Schematic plot of how a jump in $\omega_r$ results in a plateau in $\chi'$. Dashed blue curves indicates contributions from each mode, $q_1$, $q_2$, $q_3$... Solid blue curve denotes $\chi'$, which is the sum of all modes contribution. (a) If $\omega_r$ for adjacent modes are close to each other, there is no plateau. (b) A big jump between $\omega_r(q_2)$ and $\omega_r(q_3)$ results in a plateau.   }
	\label{plateau}
\end{figure}
To show the relation between a jump in the mode-dependent mobility and the plateau, in Fig.~(\ref{plateau}) we plot a discrete version of Eq.~(\ref{S32}), i.e.,~instead of a continuous $q$, we have discrete bending modes $q=n\pi/\ell$.  When plotting contributions from all bending modes together, we find that there will be a plateau only when there is a large difference between $\omega_r$ for two adjacent modes, e.g.,~$\omega_r(q_3)\gg\omega_r(q_2)$, and the plateau exists for frequencies $\omega_r(q_2)<\omega<\omega_r(q_3)$. A physical understanding for the appearance of this plateau is that, at the time scale $1/\omega_r(q_3)\ll t\ll 1/\omega_r(q_2)$, all  bending modes with $q>q_3$ are completely relaxed, thus contributing a constant pure elastic term to $\chi$, while  bending modes with $q<q_2$ are not at all relaxed, and their contribution to $\chi$ can therefore be neglected. This large difference in $\omega_r$  corresponds to a jump in $M_{qq}$, since $\omega_r=2\kappa M_{qq} q^4$. 

Now that we have an intuitive understanding of the plateau origin, let us provide a rigorous proof that a discontinuous $M_{qq}$ (or discontinuous $\omega_r(q)$) is necessary for its appearance.  For continuous $q$, we define a jump in the mode-dependent mobility for $q\in [q_1,q_2]$ through the mode-dependent relaxation rate:
\begin{equation}
\begin{aligned}
\frac{\log[\omega_r(q_2)/\omega_r(q_1)]}{\log(q_2/q_1)}\gg 1.
\end{aligned}
\label{S34}
\end{equation} 

To prove that a plateau in $\chi'$ appears if and only if the mode-dependent mobility is discontinuous we prove two propositions: (i) if there is a big jump in $\omega_r(q)$ within a narrow region of $q$, there must be a plateau in $\chi'(\omega)$ and (ii) if there is no jump in $\omega_r(q)$, there cannot be a plateau in $\chi'(\omega)$. 
\\

{\it Proof for proposition (i): if there is a big jump in $\omega_r(q)$ within a narrow region of $q$, there must be a plateau in $\chi'(\omega)$. }\\
Let the jump take place in $q\in[q_0/(1+a),(1+a)q_0]$ such that $\omega_r[q_0/(1+a)]=\omega_0/b$, $\omega_r[(1+a)q_0]=b\omega_0$. Here $a\ll1$, corresponding to a narrow region of $q$, and $b\gg1$, corresponding to a big jump in $\omega_r(q)$.  Physically, $M_{qq}$ must increase monotonically with $q$, leading to the following inequalities:
\begin{equation}
\begin{cases}
\omega_r(q)<\frac{1}{b} (\frac{q}{q_0/(1+a)})^4\omega_0 & q < q_0/(1+a)\\
\frac{1}{b} \omega_0 \leq \omega_r(q) \leq b \omega_0 & q_0/(1+a)\leq q \leq (1+a)q_0\\
\omega_r(q)>b(\frac{q}{(1+a)q_0})^4\omega_0  & q > (1+a) q_0. 
\end{cases}
\label{S35}
\end{equation}
Let us provide a (non-strict) bound for the value of $f(\omega)$ at $\omega=\omega_0$. From Eq.~(\ref{S32}), we have
\begin{align}
\chi'(\omega_0) & =C \int_0^{\infty} \frac{\omega_r^2}{\omega_r^2 + \omega_0^2 } \frac{d q}{q^4} \notag\\&\geq C \int_{(1+a)q_0}^{\infty} \frac{1}{2} \frac{d q}{q^4}\approx \frac{C}{6q_0^3}\,. \label{S29}
\end{align}
Taking the derivative of Eq.~(\ref{S32}) with respect to $\omega$ gives
\begin{align}
&\Big| \omega\frac{d \chi'}{d \omega}\Big|_{\omega=\omega_0}  = 2C \int \frac{\omega_0^2 \omega_r^2}{(\omega_r^2 + \omega_0^2)^2} \frac{d q}{q^4}\notag\\
&  \leq 2C\int^{q_0/(1+a)}_{0} \frac{ (\frac{\omega_0}{b})^2(\frac{(1+a)q}{q_0})^8}{\omega_0^2} \frac{d q}{q^4} + 2C\int_{q_0/(1+a)}^{(1+a) q_0} \frac{1}{4} \frac{d q}{q^4}\notag\\&+2C\int_{(1+a)q_0}^{\infty} \frac{\omega_0^2 }{(b\omega_0)^2(\frac{q}{(1+a) q_0})^8 } \frac{d q}{q^4}\notag\\
&\approx \frac{32C}{55b ^2q_0^3}. \label{S30}
\end{align}
We then have from Eqs.~(\ref{S29}-\ref{S30})
\begin{equation}
\begin{aligned}
|f(\omega_0)| & \leq \frac{192}{55b^2}. 
\end{aligned}
\label{S31}
\end{equation}
Since $b\gg1$, Eq.~(\ref{S31}) proves the existence of a plateau in $\chi(\omega)$. 
\\

{\it Proof for proposition (ii): if there is no jump in $\omega_r(q)$, there cannot be a plateau in $\chi'(\omega)$. }
\\
To quantify the no-jump condition, we introduce a function, 
\begin{equation}
g(q)=\frac{\diff \log(\omega_r)}{\diff \log(q)},
\label{S36}
\end{equation}
which is the local scaling exponent of $\omega_r(q)$ (assuming it is differentiable).  Since there is no jump for any interval $[q_1,q_2]$, the value of $g(q)$ must be bounded by some finite number $\alpha$, i.e. $|g(q)|\leq \alpha $ for any $q$. 

If $\chi'(\omega)$ has no plateau, $|f(\omega)|$  is not small for any $\omega$. Let $q_r(\omega)$ be the mode with relaxation rate $\omega$, i.e.,~$\omega_r(q_r(\omega))=\omega$. Since $M_{qq}$ must increase monotonically with $q$, and because $|g(q)|\leq \alpha$, we have the following inequalities:
\begin{equation}
\begin{cases}
\omega (\frac{q}{q_r(\omega)})^4 \leq \omega_r(q) \leq \omega (\frac{q}{q_r(\omega)})^\alpha& (q >q_r(\omega))\\
\omega(\frac{q}{q_r(\omega)})^4 \geq \omega_r(q)\geq\omega (\frac{q}{q_r(\omega)})^\alpha  & (q \leq q_r(\omega))\,.
\end{cases}
\label{S26}
\end{equation}
Then, from Eq.~(\ref{S32}) we have
\begin{align}
\chi' & 
 =C \int^{q_r(\omega)}_{0} \frac{\omega_r^2}{\omega_r^2 + \omega^2} \frac{d q}{q^4} + C\int_{q_r(\omega)}^{\infty} \frac{\omega_r^2}{\omega_r^2 + \omega^2} \frac{d q}{q^4}\notag\\
& \leq C\int^{q_r(\omega)}_0 \frac{\omega^2 \left(\frac{q}{q_r(\omega)}\right)^8}{\omega^2}\frac{d q}{q^4} + C\int^{\infty}_{q_r(\omega)}\frac{d q}{q^4}\notag\\
& = \frac{8C}{15 q_r(\omega)^3},\label{S24}
\end{align}
and
\begin{align}
&\!\!\Big| \omega\frac{d \chi'}{d \omega}\Big| 
\!\!= \!2C\!\left[\int^{q_r(\omega)}_{0} \frac{\omega^2 \omega_r^2}{(\omega_r^2 + \omega^2)^2} \frac{d q}{q^4} + \!\!\int_{q_r(\omega)}^{\infty} \frac{\omega^2 \omega_r^2}{(\omega_r^2 + \omega^2)^2} \frac{d q}{q^4}\right]\notag\\
& \geq 2C\int^{q_r(\omega)}_0 \frac{\omega^2(\frac{q}{q_r(\omega)})^{2\alpha}}{4\omega^2}\frac{dq}{q^4}  + 2C\int^{\infty}_{q_r(\omega)} \frac{\omega^2}{4\omega^2 (\frac{q}{q_r(\omega)})^{2\alpha}}\frac{dq}{q^4}\notag\\
& = \frac{C}{2(2\alpha - 3)  q_r^3(\omega)} + \frac{C }{2(2 \alpha + 3) q^3_r(\omega)}.\label{S25}
\end{align}
Finally using Eqs.~(\ref{S24}-\ref{S25}), we find the bound for $f(\omega)$ to be:
\begin{equation}
\begin{aligned}
\Big|f(\omega)\Big| &=\frac{1}{\chi'}\Big|\omega\frac{d \chi'}{d \omega}\Big|
&\geq\frac{1}{(2\alpha+3)}\,.
\end{aligned}
\label{S28}
\end{equation}
For finite $\alpha$, Eq.~(\ref{S28}) shows that for any value of $\omega$, $|f(\omega)|$ is not small, which means there is no plateau in $\chi'(\omega)$, and the proposition is proved. Although in the proof we assume $\omega_r(q)$ is differentiable, the proof applies to undifferentiable $\omega_r(q)$, as long as Eq.~(\ref{S26}) is satisfied. 

In this appendix we have proved that a discontinuous $M_{qq}$ is necessary in order for a plateau in $G'(\omega)$ to appear. Above we have proved two propositions to demonstrate the importance of a jump in $\omega_r(q)$ or $M_{qq}$ when there is a plateau in $\chi'(\omega)$ or $G'(\omega)$. 
\bibliographystyle{rsc} 
\bibliography{citation}

\providecommand*{\mcitethebibliography}{\thebibliography}
\csname @ifundefined\endcsname{endmcitethebibliography}
{\let\endmcitethebibliography\endthebibliography}{}
\begin{mcitethebibliography}{44}
\providecommand*{\natexlab}[1]{#1}
\providecommand*{\mciteSetBstSublistMode}[1]{}
\providecommand*{\mciteSetBstMaxWidthForm}[2]{}
\providecommand*{\mciteBstWouldAddEndPuncttrue}
  {\def\EndOfBibitem{\unskip.}}
\providecommand*{\mciteBstWouldAddEndPunctfalse}
  {\let\EndOfBibitem\relax}
\providecommand*{\mciteSetBstMidEndSepPunct}[3]{}
\providecommand*{\mciteSetBstSublistLabelBeginEnd}[3]{}
\providecommand*{\EndOfBibitem}{}
\mciteSetBstSublistMode{f}
\mciteSetBstMaxWidthForm{subitem}
{(\emph{\alph{mcitesubitemcount}})}
\mciteSetBstSublistLabelBeginEnd{\mcitemaxwidthsubitemform\space}
{\relax}{\relax}

\bibitem[Trepat \emph{et~al.}(2007)Trepat, Deng, An, Navajas, Tschumperlin,
  Gerthoffer, Butler, and Fredberg]{Trepat2007592}
X.~Trepat, L.~Deng, S.~S. An, D.~Navajas, D.~J. Tschumperlin, W.~T. Gerthoffer,
  J.~P. Butler and J.~J. Fredberg, \emph{Nature}, 2007, \textbf{447},
  592--595\relax
\mciteBstWouldAddEndPuncttrue
\mciteSetBstMidEndSepPunct{\mcitedefaultmidpunct}
{\mcitedefaultendpunct}{\mcitedefaultseppunct}\relax
\EndOfBibitem
\bibitem[Chaudhuri \emph{et~al.}(2007)Chaudhuri, Parekh, and
  Fletcher]{Chaudhuri2007}
O.~Chaudhuri, S.~H. Parekh and D.~A. Fletcher, \emph{Nature}, 2007,
  \textbf{445}, 295--298\relax
\mciteBstWouldAddEndPuncttrue
\mciteSetBstMidEndSepPunct{\mcitedefaultmidpunct}
{\mcitedefaultendpunct}{\mcitedefaultseppunct}\relax
\EndOfBibitem
\bibitem[Bonakdar \emph{et~al.}(2016)Bonakdar, Gerum, Kuhn, Sp\"{o}rrer,
  Lippert, Schneider, Aifantis, and Fabry]{Bonakdar2016}
N.~Bonakdar, R.~Gerum, M.~Kuhn, M.~Sp\"{o}rrer, A.~Lippert, W.~Schneider, K.~E.
  Aifantis and B.~Fabry, \emph{Nature Materials}, 2016, \textbf{15},
  1090--1094\relax
\mciteBstWouldAddEndPuncttrue
\mciteSetBstMidEndSepPunct{\mcitedefaultmidpunct}
{\mcitedefaultendpunct}{\mcitedefaultseppunct}\relax
\EndOfBibitem
\bibitem[Kimpton \emph{et~al.}(2015)Kimpton, Whiteley, Waters, and
  Oliver]{Kimpton2015}
L.~S. Kimpton, J.~P. Whiteley, S.~L. Waters and J.~M. Oliver, \emph{Journal of
  Mathematical Biology}, 2015, \textbf{70}, 133--171\relax
\mciteBstWouldAddEndPuncttrue
\mciteSetBstMidEndSepPunct{\mcitedefaultmidpunct}
{\mcitedefaultendpunct}{\mcitedefaultseppunct}\relax
\EndOfBibitem
\bibitem[Fabry \emph{et~al.}(2001)Fabry, Maksym, Butler, Glogauer, Navajas, and
  Fredberg]{Fabry2001}
B.~Fabry, G.~N. Maksym, J.~P. Butler, M.~Glogauer, D.~Navajas and J.~J.
  Fredberg, \emph{Physical Review Letters}, 2001, \textbf{87}, 148102\relax
\mciteBstWouldAddEndPuncttrue
\mciteSetBstMidEndSepPunct{\mcitedefaultmidpunct}
{\mcitedefaultendpunct}{\mcitedefaultseppunct}\relax
\EndOfBibitem
\bibitem[Sollich(1998)]{Sollich1998}
P.~Sollich, \emph{Phys. Rev. E}, 1998, \textbf{58}, 738--759\relax
\mciteBstWouldAddEndPuncttrue
\mciteSetBstMidEndSepPunct{\mcitedefaultmidpunct}
{\mcitedefaultendpunct}{\mcitedefaultseppunct}\relax
\EndOfBibitem
\bibitem[Bursac \emph{et~al.}(2005)Bursac, Lenormand, Fabry, Oliver, Weitz,
  Viasnoff, Butler, and Fredberg]{Bursac2005557}
P.~Bursac, G.~Lenormand, B.~Fabry, M.~Oliver, D.~A. Weitz, V.~Viasnoff, J.~P.
  Butler and J.~J. Fredberg, \emph{Nature Materials}, 2005, \textbf{4},
  557--561\relax
\mciteBstWouldAddEndPuncttrue
\mciteSetBstMidEndSepPunct{\mcitedefaultmidpunct}
{\mcitedefaultendpunct}{\mcitedefaultseppunct}\relax
\EndOfBibitem
\bibitem[Hall(1998)]{Hall1998}
A.~Hall, \emph{Science}, 1998, \textbf{279}, 509--514\relax
\mciteBstWouldAddEndPuncttrue
\mciteSetBstMidEndSepPunct{\mcitedefaultmidpunct}
{\mcitedefaultendpunct}{\mcitedefaultseppunct}\relax
\EndOfBibitem
\bibitem[Schiffhauer \emph{et~al.}(2016)Schiffhauer, Luo, Mohan, Srivastava,
  Qian, Griffis, Iglesias, and Robinson]{Schiffhauer20161473}
E.~S. Schiffhauer, T.~Luo, K.~Mohan, V.~Srivastava, X.~Qian, E.~R. Griffis,
  P.~A. Iglesias and D.~N. Robinson, \emph{Current Biology}, 2016, \textbf{26},
  1473--1479\relax
\mciteBstWouldAddEndPuncttrue
\mciteSetBstMidEndSepPunct{\mcitedefaultmidpunct}
{\mcitedefaultendpunct}{\mcitedefaultseppunct}\relax
\EndOfBibitem
\bibitem[Volkmer~Ward \emph{et~al.}(2008)Volkmer~Ward, Weins, Pollak, and
  Weitz]{Ward20084915}
S.~M. Volkmer~Ward, A.~Weins, M.~R. Pollak and D.~A. Weitz, \emph{Biophysical
  Journal}, 2008, \textbf{95}, 4915--4923\relax
\mciteBstWouldAddEndPuncttrue
\mciteSetBstMidEndSepPunct{\mcitedefaultmidpunct}
{\mcitedefaultendpunct}{\mcitedefaultseppunct}\relax
\EndOfBibitem
\bibitem[Lieleg \emph{et~al.}(2008)Lieleg, Claessens, Luan, and
  Bausch]{Lieleg2008}
O.~Lieleg, M.~M. A.~E. Claessens, Y.~Luan and A.~R. Bausch, \emph{Phys. Rev.
  Lett.}, 2008, \textbf{101}, 108101\relax
\mciteBstWouldAddEndPuncttrue
\mciteSetBstMidEndSepPunct{\mcitedefaultmidpunct}
{\mcitedefaultendpunct}{\mcitedefaultseppunct}\relax
\EndOfBibitem
\bibitem[Lieleg \emph{et~al.}(2009)Lieleg, Schmoller, Claessens, and
  Bausch]{Lieleg20094725}
O.~Lieleg, K.~M. Schmoller, M.~M.~A.~E. Claessens and A.~R. Bausch,
  \emph{Biophysical Journal}, 2009, \textbf{96}, 4725--4732\relax
\mciteBstWouldAddEndPuncttrue
\mciteSetBstMidEndSepPunct{\mcitedefaultmidpunct}
{\mcitedefaultendpunct}{\mcitedefaultseppunct}\relax
\EndOfBibitem
\bibitem[Broedersz \emph{et~al.}(2010)Broedersz, Depken, Yao, Pollak, Weitz,
  and MacKintosh]{Broedersz2010}
C.~P. Broedersz, M.~Depken, N.~Y. Yao, M.~R. Pollak, D.~A. Weitz and F.~C.
  MacKintosh, \emph{Physical Review Letters}, 2010, \textbf{105}, 238101\relax
\mciteBstWouldAddEndPuncttrue
\mciteSetBstMidEndSepPunct{\mcitedefaultmidpunct}
{\mcitedefaultendpunct}{\mcitedefaultseppunct}\relax
\EndOfBibitem
\bibitem[Yao \emph{et~al.}(2013)Yao, Broedersz, Depken, Becker, Pollak,
  MacKintosh, and Weitz]{Yao2013}
N.~Y. Yao, C.~P. Broedersz, M.~Depken, D.~J. Becker, M.~R. Pollak, F.~C.
  MacKintosh and D.~A. Weitz, \emph{Physical Review Letters}, 2013,
  \textbf{110}, 018103\relax
\mciteBstWouldAddEndPuncttrue
\mciteSetBstMidEndSepPunct{\mcitedefaultmidpunct}
{\mcitedefaultendpunct}{\mcitedefaultseppunct}\relax
\EndOfBibitem
\bibitem[M\"uller \emph{et~al.}(2014)M\"uller, Bruinsma, Lieleg, Bausch, Wall,
  and Levine]{Muller2014}
K.~W. M\"uller, R.~F. Bruinsma, O.~Lieleg, A.~R. Bausch, W.~A. Wall and A.~J.
  Levine, \emph{Phys. Rev. Lett.}, 2014, \textbf{112}, 238102\relax
\mciteBstWouldAddEndPuncttrue
\mciteSetBstMidEndSepPunct{\mcitedefaultmidpunct}
{\mcitedefaultendpunct}{\mcitedefaultseppunct}\relax
\EndOfBibitem
\bibitem[Rouse~Jr.(1953)]{RouseJr.19531272}
P.~E. Rouse~Jr., \emph{The Journal of Chemical Physics}, 1953, \textbf{21},
  1272--1280\relax
\mciteBstWouldAddEndPuncttrue
\mciteSetBstMidEndSepPunct{\mcitedefaultmidpunct}
{\mcitedefaultendpunct}{\mcitedefaultseppunct}\relax
\EndOfBibitem
\bibitem[de~Gennes(1990)]{gennes_1990}
P.~G. de~Gennes, \emph{Introduction to Polymer Dynamics}, Cambridge University
  Press, 1990\relax
\mciteBstWouldAddEndPuncttrue
\mciteSetBstMidEndSepPunct{\mcitedefaultmidpunct}
{\mcitedefaultendpunct}{\mcitedefaultseppunct}\relax
\EndOfBibitem
\bibitem[Doi and Edwards(1988)]{doi}
M.~Doi and S.~F. Edwards, \emph{The Theory of Polymer Dynamics}, Clarendon
  Press, 1988\relax
\mciteBstWouldAddEndPuncttrue
\mciteSetBstMidEndSepPunct{\mcitedefaultmidpunct}
{\mcitedefaultendpunct}{\mcitedefaultseppunct}\relax
\EndOfBibitem
\bibitem[Gittes \emph{et~al.}(1997)Gittes, Schnurr, Olmsted, MacKintosh, and
  Schmidt]{Gittes1997}
F.~Gittes, B.~Schnurr, P.~D. Olmsted, F.~C. MacKintosh and C.~F. Schmidt,
  \emph{Phys. Rev. Lett.}, 1997, \textbf{79}, 3286--3289\relax
\mciteBstWouldAddEndPuncttrue
\mciteSetBstMidEndSepPunct{\mcitedefaultmidpunct}
{\mcitedefaultendpunct}{\mcitedefaultseppunct}\relax
\EndOfBibitem
\bibitem[Gittes and MacKintosh(1998)]{Gittes1998}
F.~Gittes and F.~C. MacKintosh, \emph{Physical Review E}, 1998, \textbf{58},
  R1241--R1244\relax
\mciteBstWouldAddEndPuncttrue
\mciteSetBstMidEndSepPunct{\mcitedefaultmidpunct}
{\mcitedefaultendpunct}{\mcitedefaultseppunct}\relax
\EndOfBibitem
\bibitem[Morse(1998)]{Morse1998}
D.~C. Morse, \emph{Phys. Rev. E}, 1998, \textbf{58}, R1237--R1240\relax
\mciteBstWouldAddEndPuncttrue
\mciteSetBstMidEndSepPunct{\mcitedefaultmidpunct}
{\mcitedefaultendpunct}{\mcitedefaultseppunct}\relax
\EndOfBibitem
\bibitem[Morse(1998)]{Morse19987030}
D.~C. Morse, \emph{Macromolecules}, 1998, \textbf{31}, 7030--7043\relax
\mciteBstWouldAddEndPuncttrue
\mciteSetBstMidEndSepPunct{\mcitedefaultmidpunct}
{\mcitedefaultendpunct}{\mcitedefaultseppunct}\relax
\EndOfBibitem
\bibitem[Koenderink \emph{et~al.}(2006)Koenderink, Atakhorrami, MacKintosh, and
  Schmidt]{Koenderink2006}
G.~H. Koenderink, M.~Atakhorrami, F.~C. MacKintosh and C.~F. Schmidt,
  \emph{Phys. Rev. Lett.}, 2006, \textbf{96}, 138307\relax
\mciteBstWouldAddEndPuncttrue
\mciteSetBstMidEndSepPunct{\mcitedefaultmidpunct}
{\mcitedefaultendpunct}{\mcitedefaultseppunct}\relax
\EndOfBibitem
\bibitem[Broedersz and MacKintosh(2014)]{Broedersz2014}
C.~P. Broedersz and F.~C. MacKintosh, \emph{Reviews of Modern Physics}, 2014,
  \textbf{86}, 995--1036\relax
\mciteBstWouldAddEndPuncttrue
\mciteSetBstMidEndSepPunct{\mcitedefaultmidpunct}
{\mcitedefaultendpunct}{\mcitedefaultseppunct}\relax
\EndOfBibitem
\bibitem[Pritchard \emph{et~al.}(2014)Pritchard, Shery~Huang, and
  Terentjev]{Pritchard2014}
R.~Pritchard, Y.~Shery~Huang and E.~Terentjev, \emph{Soft Matter}, 2014,
  \textbf{10}, 1864--1884\relax
\mciteBstWouldAddEndPuncttrue
\mciteSetBstMidEndSepPunct{\mcitedefaultmidpunct}
{\mcitedefaultendpunct}{\mcitedefaultseppunct}\relax
\EndOfBibitem
\bibitem[Alberts \emph{et~al.}(2017)Alberts, Johnson, Lewis, Morgan, Raff,
  Roberts, and Walter]{Alberts}
B.~Alberts, A.~D. Johnson, J.~Lewis, D.~Morgan, M.~Raff, K.~Roberts and
  P.~Walter, \emph{Molecular Biology of the Cell}, Garland Science, New York,
  6th edn., 2017\relax
\mciteBstWouldAddEndPuncttrue
\mciteSetBstMidEndSepPunct{\mcitedefaultmidpunct}
{\mcitedefaultendpunct}{\mcitedefaultseppunct}\relax
\EndOfBibitem
\bibitem[Bray(2001)]{Bray2001}
D.~Bray, \emph{Cell Movements: From Molecules to Motility}, Garland, 2nd edn.,
  2001\relax
\mciteBstWouldAddEndPuncttrue
\mciteSetBstMidEndSepPunct{\mcitedefaultmidpunct}
{\mcitedefaultendpunct}{\mcitedefaultseppunct}\relax
\EndOfBibitem
\bibitem[Markovich \emph{et~al.}(2019)Markovich, Tjhung, and
  Cates]{Markovich2019}
T.~Markovich, E.~Tjhung and M.~E. Cates, \emph{New Journal of Physics}, 2019,
  \textbf{21}, year\relax
\mciteBstWouldAddEndPuncttrue
\mciteSetBstMidEndSepPunct{\mcitedefaultmidpunct}
{\mcitedefaultendpunct}{\mcitedefaultseppunct}\relax
\EndOfBibitem
\bibitem[Wang \emph{et~al.}(2002)Wang, Toli{\'c}-N\o{}rrelykke, Chen,
  Mijailovich, Butler, Fredberg, and Stamenovi{\'c}]{Wang2002}
N.~Wang, I.~M. Toli{\'c}-N\o{}rrelykke, J.~Chen, S.~M. Mijailovich, J.~P.
  Butler, J.~J. Fredberg and D.~Stamenovi{\'c}, \emph{American Journal of
  Physiology - Cell Physiology}, 2002, \textbf{282}, C606--C616\relax
\mciteBstWouldAddEndPuncttrue
\mciteSetBstMidEndSepPunct{\mcitedefaultmidpunct}
{\mcitedefaultendpunct}{\mcitedefaultseppunct}\relax
\EndOfBibitem
\bibitem[Stamenovi{\'{c}} \emph{et~al.}(2004)Stamenovi{\'{c}}, Suki, Fabry,
  Wang, and Fredberg]{Stamenovic2004}
D.~Stamenovi{\'{c}}, B.~Suki, B.~Fabry, N.~Wang and J.~Fredberg, \emph{Journal
  of Applied Physiology}, 2004, \textbf{96}, 1600--1605\relax
\mciteBstWouldAddEndPuncttrue
\mciteSetBstMidEndSepPunct{\mcitedefaultmidpunct}
{\mcitedefaultendpunct}{\mcitedefaultseppunct}\relax
\EndOfBibitem
\bibitem[Fern{\'a}ndez \emph{et~al.}(2006)Fern{\'a}ndez, Pullarkat, and
  Ott]{FERNANDEZ2006}
P.~Fern{\'a}ndez, P.~A. Pullarkat and A.~Ott, \emph{Biophysical Journal}, 2006,
  \textbf{90}, 3796--3805\relax
\mciteBstWouldAddEndPuncttrue
\mciteSetBstMidEndSepPunct{\mcitedefaultmidpunct}
{\mcitedefaultendpunct}{\mcitedefaultseppunct}\relax
\EndOfBibitem
\bibitem[Mulla \emph{et~al.}(2019)Mulla, MacKintosh, and Koenderink]{Mulla2019}
Y.~Mulla, F.~C. MacKintosh and G.~H. Koenderink, \emph{Physical Review
  Letters}, 2019, \textbf{122}, 218102\relax
\mciteBstWouldAddEndPuncttrue
\mciteSetBstMidEndSepPunct{\mcitedefaultmidpunct}
{\mcitedefaultendpunct}{\mcitedefaultseppunct}\relax
\EndOfBibitem
\bibitem[J{\"{u}}licher \emph{et~al.}(2007)J{\"{u}}licher, Kruse, Prost, and
  Joanny]{Julicher20073}
F.~J{\"{u}}licher, K.~Kruse, J.~Prost and J.-F. Joanny, \emph{Physics Reports},
  2007, \textbf{449}, 3--28\relax
\mciteBstWouldAddEndPuncttrue
\mciteSetBstMidEndSepPunct{\mcitedefaultmidpunct}
{\mcitedefaultendpunct}{\mcitedefaultseppunct}\relax
\EndOfBibitem
\bibitem[Roh-Johnson \emph{et~al.}(2012)Roh-Johnson, Shemer, Higgins,
  McClellan, Werts, Tulu, Gao, Betzig, Kiehart, and
  Goldstein]{Roh-Johnson20121232}
M.~Roh-Johnson, G.~Shemer, C.~D. Higgins, J.~H. McClellan, A.~D. Werts, U.~S.
  Tulu, L.~Gao, E.~Betzig, D.~P. Kiehart and B.~Goldstein, \emph{Science},
  2012, \textbf{335}, 1232--1235\relax
\mciteBstWouldAddEndPuncttrue
\mciteSetBstMidEndSepPunct{\mcitedefaultmidpunct}
{\mcitedefaultendpunct}{\mcitedefaultseppunct}\relax
\EndOfBibitem
\bibitem[Chen \emph{et~al.}(2020)Chen, Markovich, and MacKintosh]{Chen2020}
S.~Chen, T.~Markovich and F.~C. MacKintosh, \emph{Phys. Rev. Lett.}, 2020,
  \textbf{125}, 208101\relax
\mciteBstWouldAddEndPuncttrue
\mciteSetBstMidEndSepPunct{\mcitedefaultmidpunct}
{\mcitedefaultendpunct}{\mcitedefaultseppunct}\relax
\EndOfBibitem
\bibitem[Hohenberg and Halperin(1977)]{Hohenberg1977435}
P.~C. Hohenberg and B.~I. Halperin, \emph{Reviews of Modern Physics}, 1977,
  \textbf{49}, 435--479\relax
\mciteBstWouldAddEndPuncttrue
\mciteSetBstMidEndSepPunct{\mcitedefaultmidpunct}
{\mcitedefaultendpunct}{\mcitedefaultseppunct}\relax
\EndOfBibitem
\bibitem[Chen \emph{et~al.}()Chen, Markovich, and MacKintosh]{unpublished}
S.~Chen, T.~Markovich and F.~C. MacKintosh, \emph{unpublished}\relax
\mciteBstWouldAddEndPuncttrue
\mciteSetBstMidEndSepPunct{\mcitedefaultmidpunct}
{\mcitedefaultendpunct}{\mcitedefaultseppunct}\relax
\EndOfBibitem
\bibitem[Note1()]{Note1}
The collapse of the experimental data was previously done in Ref.~{\cite
  {Mulla2019}} using a phenomenological theory.\relax
\mciteBstWouldAddEndPunctfalse
\mciteSetBstMidEndSepPunct{\mcitedefaultmidpunct}
{}{\mcitedefaultseppunct}\relax
\EndOfBibitem
\bibitem[Velankar and Giles(2007)]{Velankar2007}
S.~S. Velankar and D.~Giles, \emph{The News and Information Publication of The
  Society of Rheology}, 2007, \textbf{76}, 8--20\relax
\mciteBstWouldAddEndPuncttrue
\mciteSetBstMidEndSepPunct{\mcitedefaultmidpunct}
{\mcitedefaultendpunct}{\mcitedefaultseppunct}\relax
\EndOfBibitem
\bibitem[Kollmannsberger \emph{et~al.}(2011)Kollmannsberger, Mierke, and
  Fabry]{Kollmannsberger20113127}
P.~Kollmannsberger, C.~T. Mierke and B.~Fabry, \emph{Soft Matter}, 2011,
  \textbf{7}, 3127--3132\relax
\mciteBstWouldAddEndPuncttrue
\mciteSetBstMidEndSepPunct{\mcitedefaultmidpunct}
{\mcitedefaultendpunct}{\mcitedefaultseppunct}\relax
\EndOfBibitem
\bibitem[Desprat \emph{et~al.}(2005)Desprat, Richert, Simeon, and
  Asnacios]{Desprat20052224}
N.~Desprat, A.~Richert, J.~Simeon and A.~Asnacios, \emph{Biophysical Journal},
  2005, \textbf{88}, 2224--2233\relax
\mciteBstWouldAddEndPuncttrue
\mciteSetBstMidEndSepPunct{\mcitedefaultmidpunct}
{\mcitedefaultendpunct}{\mcitedefaultseppunct}\relax
\EndOfBibitem
\bibitem[Balland \emph{et~al.}()Balland, Desprat, Icard, F\'er\'eol, Asnacios,
  Browaeys, H\'enon, and Gallet]{Balland2006}
M.~Balland, N.~Desprat, D.~Icard, S.~F\'er\'eol, A.~Asnacios, J.~Browaeys,
  S.~H\'enon and F.~Gallet\relax
\mciteBstWouldAddEndPuncttrue
\mciteSetBstMidEndSepPunct{\mcitedefaultmidpunct}
{\mcitedefaultendpunct}{\mcitedefaultseppunct}\relax
\EndOfBibitem
\bibitem[Navarro \emph{et~al.}(1996)Navarro, Frenk, and White]{Navarro1996}
J.~F. Navarro, C.~S. Frenk and S.~D.~M. White, 1996, \textbf{1}, 493--508\relax
\mciteBstWouldAddEndPuncttrue
\mciteSetBstMidEndSepPunct{\mcitedefaultmidpunct}
{\mcitedefaultendpunct}{\mcitedefaultseppunct}\relax
\EndOfBibitem
\bibitem[Cloitre \emph{et~al.}(2003)Cloitre, Borrega, Monti, and
  Leibler]{Cloitre2003}
M.~Cloitre, R.~Borrega, F.~Monti and L.~Leibler, \emph{Physical Review
  Letters}, 2003, \textbf{90}, 068303\relax
\mciteBstWouldAddEndPuncttrue
\mciteSetBstMidEndSepPunct{\mcitedefaultmidpunct}
{\mcitedefaultendpunct}{\mcitedefaultseppunct}\relax
\EndOfBibitem
\end{mcitethebibliography}
\end{document}